\documentclass[fleqn,usenatbib]{mnras}

\usepackage{newtxtext,newtxmath}
\usepackage[T1]{fontenc}

\DeclareRobustCommand{\VAN}[3]{#2}
\let\VANthebibliography\thebibliography
\def\thebibliography{\DeclareRobustCommand{\VAN}[3]{##3}\VANthebibliography}

\usepackage{graphicx}	% Including figure files
\usepackage{amsmath}	% Advanced maths commands
\usepackage{nccmath}

\usepackage{booktabs}
\usepackage{chemformula}
\usepackage[capitalise]{cleveref}

\title[\ch{H2} Thermal Dissociation/Recombination in UHJs]{Pseudo-2D Modelling of Heat Redistribution Through \ch{H2} Thermal Dissociation/Recombination: Consequences for Ultra-Hot Jupiters}

\author[A. Roth et al.]{
A. Roth,$^{1,2}$\thanks{alexander.roth@physics.ox.ac.uk}
B. Drummond,$^{3,2}$
E. H\'ebrard,$^{2}$
P. Tremblin,$^{4}$
J. Goyal$^{5}$
and N. Mayne$^{2}$
\\
$^{1}$Department of Physics (Atmospheric, Oceanic and Planetary Physics), University of Oxford, Parks Rd, Oxford, OX1 3PU, UK\\
$^{2}$Astrophysics Group, University of Exeter, Stocker Road, Exeter, EX4 4QL, UK\\
$^{3}$Met Office, Fitzroy Road, Exeter, EX1 3PB, UK\\
$^{4}$Université Paris-Saclay, UVSQ, CNRS, CEA, Maison de la Simulation, 91191, Gif-sur-Yvette, FRANCE\\
$^{5}$Department of Astronomy and Carl Sagan Institute, Cornell University, 122 Sciences Drive, Ithaca, NY, 14853, USA
}

\date{Accepted XXX. Received YYY; in original form ZZZ}

\pubyear{2020}

\begin{document}
\label{firstpage}
\pagerange{\pageref{firstpage}--\pageref{lastpage}}
\maketitle

% Abstract of the paper
\begin{abstract}
Thermal dissociation and recombination of molecular hydrogen, \ch{H2}, in the atmospheres of ultra-hot Jupiters (UHJs) has been shown to play an important role in global heat redistribution. This, in turn, significantly impacts their planetary emission, yet only limited investigations on the atmospheric effects have so far been conducted. Here we investigate the heat redistribution caused by this dissociation/recombination reaction, alongside feedback mechanisms between the atmospheric chemistry and radiative transfer, for a planetary and stellar configuration typical of UHJs. To do this, we have developed a time-dependent pseudo-2D model, including a treatment of time-independent equilibrium chemical effects. As a result of the reaction heat redistribution, we find temperature changes of up to $\sim$400 K in the atmosphere. When \ch{TiO} and \ch{VO} are additionally considered as opacity sources, these changes in temperature increase to over $\sim$800 K in some areas. This heat redistribution is found to significantly shift the region of peak atmospheric temperature, or hotspot, towards the evening terminator in both cases. The impact of varying the longitudinal wind speed on the reaction heat distribution is also investigated. When excluding \ch{TiO}/\ch{VO}, increased wind speeds are shown to increase the impact of the reaction heat redistribution up to a threshold wind speed. When including \ch{TiO}/\ch{VO} there is no apparent wind speed threshold, due to thermal stabilisation by these species. We also construct pseudo-2D phase curves from our model, and highlight both significant spectral flux damping and increased phase offset caused by the reaction heat redistribution.
\end{abstract}

% Select between one and six entries from the list of approved keywords.
% Don't make up new ones.
\begin{keywords}
   planets and satellites: atmospheres -- planets and satellites: composition -- planets and satellites: gaseous planets
\end{keywords}

%%%%%%%%%%%%%%%%%%%%%%%%%%%%%%%%%%%%%%%%%%%%%%%%%%

%%%%%%%%%%%%%%%%% BODY OF PAPER %%%%%%%%%%%%%%%%%%

\section{Introduction}
\label{sec:introduction}
Offering both valuable insight into extreme planetary atmospheric conditions, and an abundance of available data due to detection biases, the hydrogen-dominated atmospheres of highly irradiated ultra-hot Jupiter (UHJ) exoplanets have recently become a focus of both the modelling and observational astrophysics communities. These planets are broadly assumed to be tidally-locked, due to strong gravitational interactions with their host star, causing huge irradiative disparities between their permanently stellar-facing day-side and much colder night-side. Sophisticated multi-dimensional atmospheric models exist for these planets, and high speed variable winds have been predicted through modelling \citep{{2002A&A...385..156G}, {2014A&A...561A...1M}} and inferred from observations \citep{2010Natur.465.1049S}. For UHJs, such as WASP-12b, the day-side equilibrium temperature can reach $\geq$ 2500 K \citep{2019AJ....158...39C}, with night-side temperatures $\geq$ 1000 K colder \citep{2012ApJ...747...82C}. This extreme temperature gradient between the two hemispheres drives global circulation in the form of particularly strong longitudinal winds, dominated by an equatorial jet, in some cases reaching speeds of over 5 km s$^{-1}$ \citep{2017ApJ...835..198K}.

For the range of typical atmospheric pressures, temperatures between 2000 -- 4000 K cause molecular hydrogen (\ch{H2}) to thermally dissociate into atomic hydrogen (H). Energy is absorbed during the endothermic dissociation reaction (\ch{H2} $\rightarrow$ H + H) and released during the exothermic recombination reaction (H + H $\rightarrow$ \ch{H2}). \citet{2018ApJ...857L..20B} have shown that dissociation on the day-side followed by recombination on the much cooler night-side, driven by the large temperature difference between the substellar and antistellar hemispheres, reduce the overall day-night temperature gradient. This heat redistribution effect has been demonstrated to impact photometric measurements for ultra-hot Jupiters, specifically transit photometry, both theoretically and observationally \citep{Komacek_2018,2019ApJ...886...26T,Mansfield_2020}. The early work by \citet{2018ApJ...857L..20B} is based on a simple atmospheric energy balance model. In this paper we further investigate this process using a more sophisticated atmosphere model that includes equilibrium chemistry, radiative transfer and a parameterised horizontal transport.

A more thorough investigation on the consequences caused by \ch{H2} thermal dissociation/recombination must include the treatment of complex chemical feedback within the atmosphere. One-dimensional (1D) radiative transfer and chemistry models, which were built to investigate planetary temperature structures and the associated chemical composition by solving for radiative-convective equilibrium, are a typical starting point for this. Most of these models make the assumption of local chemical equilibrium \citep[e.g.,][]{Molliere_2015,Gandhi_2017}, which is likely a good assumption for very hot atmospheres. Some models have also incorporated the ability to consistently calculate the temperature profile with non-equilibrium chemistry, including vertical mixing and photochemistry \citep{2016A&A...594A..69D}. These models have since been used to investigate a range of planetary atmospheres, such as hot Neptunes \citep{2014ApJ...781...68A} and brown dwarfs \citep{2020arXiv200313717P}, but have predominantly been used for hot Jupiters \citep{{2013ApJ...763...25M}, {2020MNRAS.tmp.2424G}}.

The use of 1D models for planets like ultra-hot Jupiters, which display such extreme longitudinal atmospheric change, clearly has significant limitations. These models, by design, average incoming stellar flux to a single value, severely limiting their use for any study on effects caused by longitudinal temperature or chemical variations. Three-dimensional (3D) global circulation models can be used to model more intricate planetary dynamics \citep[e.g.][]{2016A&A...595A..36A}, and have been applied to the \ch{H2} dissociation/recombination effect \citep{2019ApJ...886...26T}. However, due to their increased  complexity and computational cost, handling of the expensive chemistry calculations is typically simplified compared with the approach in 1D models.

It therefore stands to reason that, in order to properly investigate the affects of \ch{H2} thermal dissociation/recombination, some fundamental changes and additions must be made to existing models. Both pseudo-2D \citep{2014A&A...564A..73A} and time-dependent \citep{2005A&A...436..719I} models have previously been adapted from pre-existing 1D models and used to investigate hot Jupiters. In this work we further develop an existing 1D/2D atmosphere model (ATMO) to include a time-dependent pseudo-2D capability. We then use this to study the effect of \ch{H2} thermal dissociation/recombination, and its importance within UHJ atmospheres. Among the most important effects previously unaccounted for, and a primary focus of this study, are the numerous interactions between \ch{H2} thermal dissociation/recombination and important atmospheric opacity sources, particularly \ch{TiO} and \ch{VO}, which are known to drastically affect the pressure-temperature structure within these atmospheres \citep{Fortney_2008,2018A&A...617A.110P}. 

A brief review and details of new developments to the ATMO model are presented in \cref{sec:themodel}. Our results and discussion on heat redistribution from the thermal dissociation/recombination of hydrogen can be found in \cref{sec:tdh}. A conclusion is then drawn on all the findings in \cref{sec:conc}.
%
%--------------------------------------------------------------------
\section{Model Description and Method}
\label{sec:themodel}
\subsection{Atmosphere model: ATMO}
\label{subsec:ATMO}
\begin{table}
\centering
\caption{Sources for opacity data used within ATMO}
\begin{tabular}{@{} c @{} c @{}}    % l - left justified within column, 
                                    % r - right justified within column, 
                                    % c - centered in column
\toprule 
Species & Source  \\ 
\midrule 
\ch{H2}-\ch{H2}, \ch{H2}-He & HITRAN \citep{2012JQSRT.113.1276R} \\
\ch{CH4} & \citet{2014MNRAS.440.1649Y} \\
\ch{H2O} & \citet{2006MNRAS.368.1087B} \\
\ch{CO} & \citet{2010JQSRT.111.2139R} \\ 
\ch{CO2} & \citet{2011JQSRT.112.1403T} \\
\ch{NH3} & \citet{2011MNRAS.413.1828Y} \\
%\ch{HCN} & \citet{Harris_2006,Barber_2014}\\
%\ch{C2H2} & \citet{Rothman_2013}\\
\midrule
 $\kern-\nulldelimiterspace\left.
 \begin{tabular}{@{}l@{}}
    \ch{Na} \\
    \ch{K}  \\
    \ch{Li} \\
    \ch{Rb} \\
    \ch{Cs}
  \end{tabular}\right\}$ & \citet{2015PhyS...90e4010H};\citet{Ryabchikova_2015} \\
\midrule
\ch{TiO} & \citet{Plez1998} \\
\ch{VO} & \citet{2016MNRAS.463..771M} \\
\ch{H-} & \citet{John1988}  \\
\bottomrule
\end{tabular}
\label{tab:opacity}
\end{table}

The 1D/2D atmosphere code ATMO \citep{2015ApJ...804L..17T, 2016A&A...594A..69D, 2017ApJ...841...30T, 2018MNRAS.474.5158G} has been widely used to model substellar atmospheres. Most typically the model is used in a 1D form, solving for the 1D radiative-convective equilibrium profile of an atmosphere. An extension of the model by \citet{2017ApJ...841...30T} included the ability to solve for the stationary-state 2D (longitude-altitude) atmosphere. ATMO has been applied to hydrogen-dominated, irradiated exoplanet atmospheres \citep[e.g.][]{2018MNRAS.474.5158G, 2019MNRAS.486.1123D}, brown dwarf atmospheres \citep[e.g.][]{2016ApJ...817L..19T, 2020arXiv200313717P} and as a retrieval code for constraining observations \citep[e.g.][]{Evans2017,Nikolov2018,Wakeford2018}.

In its 1D form, ATMO solves for hydrostatic balance and radiative-convective equilibrium, with an internal heat flux (we use an intrinsic temperature of $T_\text{int}$ = 100 K) and irradiation at the top of the atmosphere as boundary conditions. The radiative transfer equation is solved in 1D plane-parallel geometry and includes isotropic scattering. We include absorption by \ch{CH4}, \ch{H2O}, \ch{CO}, \ch{CO2}, \ch{NH3}, \ch{Na}, \ch{K}, \ch{Li}, \ch{Rb}, \ch{Cs}, \ch{TiO}, \ch{VO}, and \ch{H-} as opacity sources. \ch{H2}-\ch{H2} and \ch{H2}-\ch{He} collision-induced absorption is also included. All references for the opacity source data used in this study can be found in \cref{tab:opacity}. The most up-to-date description of the line-lists and calculation of the opacities can be found in \citet{2020MNRAS.tmp.2424G}.
The model uses the correlated-$k$ approximation with the method of random overlap to calculate the combined opacity of the mixture \citep{Lacis1991,AmuTM17}, with 32 bands to compute the radiative flux in the radiative-convective equilibrium iterations. 

Chemical equilibrium abundances are obtained by minimising the Gibbs free energy, following the method of \citet{GordonMcBride1994}. The thermodynamic properties of all chemical species are expressed in terms of NASA polynomials, using coefficients from \citet{Mcbride1993, McBride2002}. The Gibbs minimisation method allows for the depletion of gas phase species due to condensation, but we only consider a gas-phase composition here. This is likely to be a good assumption for the results presented in this paper, since the high-temperatures of the atmospheres that we focus on in this study mean that condensation is expected to be unimportant.

The pressure and temperature dependence of chemical equilibrium abundances creates the requirement for consistency between the radiative-convective and chemistry calculations. As the pressure-temperature ($P$ -- $T$) structure evolves, chemical abundances shift to updated values which in turn changes the opacity. This in turn affects the radiative transfer and therefore the $P$ -- $T$ structure. By periodically cycling between the radiative-convective solver and the Gibbs energy minimisation scheme, ATMO can account for this feedback mechanism and determine $P$ -- $T$ and chemical profiles that satisfy radiative-convective equilibrium.

ATMO also includes the option to model time-dependent non-equilibrium chemical effects, such as vertical mixing and photochemistry. However, for this study, we only include time-independent chemical equilibrium processes.
%
%%%%%%%%%%%%%%%%%%%%%%%%%%%%%%%%%%%%%%%%%%%%%%%%%%%%%%%%%%%%%%%
\subsection{A time-dependent approach to solving for 1D radiative equilibrium}
\label{subsec:tdmodel}
Most models, including previous applications of ATMO, use a time-independent flux-balancing approach when solving for radiative-convective equilibrium. We refer to this approach as the `time-independent' approach as it involves applying incremental perturbations to the $P$ -- $T$ profile until a profile that satisfies radiative-convective balance is achieved. An alternative approach is to calculate heating and cooling rates and increment the temperature on each pressure level until the heating rate reduces to zero, to within some tolerance, for all model levels \citep[][]{2005A&A...436..719I, 2017AJ....153...56M}. We refer to this second approach as the `time-dependent' approach, since it involves iterating through time-steps until a steady-state is found.

Within ATMO, we have recently implemented a time-dependent approach to solving for radiative equilibrium, following the same basic method outlined in previous studies \citep{2005A&A...436..719I,2017AJ....153...56M}. The governing energy equation is
\begin{ceqn}
\begin{equation}
\frac{dT}{dt}=\frac{1}{c_P \rho}\frac{dF}{dz},
\label{eq:1}
\end{equation}
\end{ceqn}
\noindent
where $T$ is temperature, $t$ is time, $c_P$ is the specific heat capacity, $\rho$ is the gas density, and $z$ is altitude. We note that Equation \cref{eq:1} can be equivalently expressed in terms of the pressure gradient, rather than the altitude gradient. The net radiative flux, $F$, is calculated using the radiative transfer scheme in ATMO \citep{2014A&A...564A..59A, 2016A&A...594A..69D}. Convective flux is not presently included, though as UHJ atmospheres are dominated by radiative energy transfer, and for the most part convectively stable, we expect convective flux to be unimportant. The $P$ -- $T$ profile that satisfies radiative equilibrum corresponds to a steady-state solution of \cref{eq:1} (i.e. $dT/dt = 0$, to within some tolerance).

The newly implemented time-dependent radiative equilibrium solver has been validated by comparing the steady-state radiative equilibrium $P$ -- $T$ profile against a $P$ -- $T$ profile calculated using the well-established, and well-tested, time-independent flux-balancing scheme within ATMO. We note that the latter has previously been compared against other 1D atmosphere models for both brown dwarfs and exoplanets \citep{Baudino2017, Malik2019}. Details of this test are presented in \cref{appendix:benchmark}.

The atmosphere is discretised onto a pressure grid that is uniformly spaced in the $\log_{10}(P)$-space. The vertical grid is then defined by the maximum and minimum pressure and the number of pressure levels. Altitude increments ($\Delta z$) between model levels are calculated assuming local hydrostatic balance.

An adaptive time-stepping method is used to handle the fact that the deep atmosphere evolves much more slowly than the upper atmosphere, where the radiative timescale is much shorter, as applied in \citet{2005A&A...436..719I} and \citet{2017AJ....153...56M}. When solving for the radiative equilibrium temperature profile, the timestep is allowed to vary independently in each model level, allowing much larger timesteps for model levels at higher pressures. This allows for convergence to be reached in a reasonable number of iterations in the deep atmosphere, while maintaining a timestep that is short enough to remain numerically stable in the upper atmosphere. Applying a different timestep across model levels is acceptable as long as the steady-state solution is sought, though it means that the profiles at intermediate steps before convergence is reached are not physically consistent.

At the start of a new calculation, the timestep in each level is initialised based on the radiative timescale,
\begin{ceqn}
\begin{equation}
\tau_{\rm rad} \sim \frac{P}{g}\frac{c_P}{\sigma T^3},
\label{eq:eq2}
\end{equation}
\end{ceqn}
\noindent
where $g$ is gravity and $\sigma$ is the Stefan-Boltzmann constant. At the top of the atmosphere (for $P=1$~Pa) the timestep is of order $\sim10$~s, while at the bottom of the atmosphere (for $P=10^7$~Pa) the timestep is typically of order $\sim10^8$~s, though these values strongly depend on temperature. Following each iteration the timestep in each model level is increased by 10\%, except if several criteria are met that are introduced to maintain model stability. If the heating rate in a model level has switched to an opposite sign compared with the previous timestep (indicating an oscillating heating rate in time) then the timestep is reduced by 33\%. Additionally, if the heating rate in a model level has the opposite sign to both of its neighbouring model levels (indicating a heating rate that is oscillating in space) then the timestep is reduced by 33\%. We note that when considering a time-dependent phenomenon (such as a varying instellation, as described later) we use a fixed and uniform timestep for all model levels instead.

We also introduce a limit on the magnitude of the temperature increments within each level. This prevents excessively large adjustments to the temperature, which can be present especially in the early stages of a calculation when starting from an initial condition far from the equilibrium state (e.g. an isothermal profile). We set this temperature increment cap to $\pm50$ K.

When solving for radiative equilibrium it is important to define a set of criterion to signal when convergence has been reached. For the steady-state we test the normalised flux gradient, which approaches zero as the profile approaches radiative equilibrium, following the approach of \citet{2017AJ....153...56M}. The criterion for convergence is that
\begin{ceqn}
\begin{equation}
\frac{\Delta F}{\sigma T^4} < \delta,
\label{eq:eq3}
\end{equation}
\end{ceqn}
\noindent
for all model levels, where $\delta$ is some set tolerance. We use a tolerance of $\delta=10^{-3}$.

In this work, the time-dependent radiative equilibrium solver described in this section is used to generate initial profiles for the pseudo-2D model described in the next section.
%
%%%%%%%%%%%%%%%%%%%%%%%%%%%%%%%%%%%%%%%%%%%%%%%%%%%%%%%%%%%%%%%
\subsection{Pseudo-2D modelling}
\label{subsec:tdmodel2}
\begin{figure*}
\centering
\includegraphics[width=\textwidth]{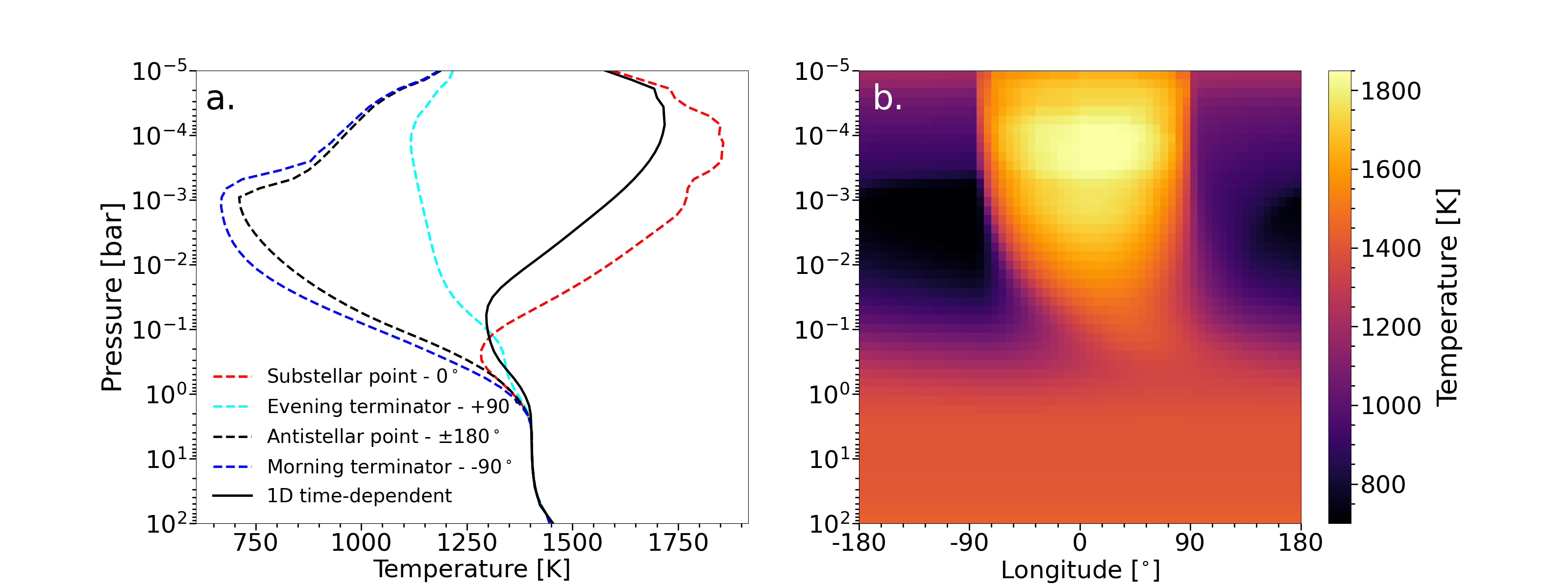}
\caption{(a) Steady-state $P$ -- $T$ profiles from the pseudo-2D model at four key longitudinal positions: the substellar point ($0^\circ$), evening terminator ($+90^\circ$), antistellar point ($\pm180^\circ$) and morning terminator ($-90^\circ$), compared with the steady-state 1D time-dependent model seen in \cref{fig:benchmark}. (b) Periodic steady-state $P$ -- $T$ structure for an HJ atmosphere as it evolves with longitudinal position, calculated with the pseudo-2D model. All opacity sources (including TiO and VO) as listed in \cref{subsec:ATMO} are included and a wind speed of $\sim$1 km s$^{-1}$ (R=600000s, giving a wind speed of 1010 m s$^{-1}$) is used.}
\label{fig:fig1}
\end{figure*}
With time-dependency introduced in ATMO, the next step is to continuously rotate the atmospheric column around the equator, creating a pseudo-2D model. This introduces a simple parameterisation of horizontal advection. We follow the method used in \citet{2005A&A...436..719I}. A time-dependent modulating function, $\alpha$, is applied to the incoming stellar flux,
\begin{ceqn}
\begin{equation}
\begin{aligned}
&F_{star,incident}^0 = \alpha F_{star}^0 , \\
&\alpha = \pi \textrm{max}\{\cos{\lambda},0\}, \quad \lambda = \frac{2\pi t}{R},
\label{eq:4}
\end{aligned}
\end{equation}
\end{ceqn}
\noindent
which adjusts incident radiation based on the zenith angle of the host star, so that radiation is at a maximum when the star is directly overhead, and zero at all points on the night-side. $t$ is the same time as found in \cref{eq:1}. $R$ is the column rotation period (not to be confused with the planetary rotation period). The value of $R$ is selected, on a case by case basis, in order to achieve the desired effective wind speed at the planetary equator. To determine if a periodic steady-state has been achieved, each subsequent full model rotation is qualitatively compared to the last, as frequently done in GCMs \citep[e.g.][]{2020A&A...636A..68D}. Qualitative comparison is often used in this way as the deep atmosphere, where the radiative and dynamical timescales are extremely long, does not reach a steady-state in a realistic time due to computational limitations \citep{Mayne2017}. Once a periodic steady-state has been achieved, the model time, $t$, is used as an analogue to the equatorial longitudinal position of the model.
\begin{table}
\centering
\caption{Planetary and stellar parameters used within ATMO. We assume a planetary albedo, $\alpha$ = 0, and emissivity, $\epsilon$ = 1, in all cases.}
\begin{tabular}{ c c c c }   % l - left justified within column, 
                                         % r - right justified within column,             
                                         % c - centered in column
\toprule 

Parameter & Unit & HD 209458b & WASP-12b \\ 
\midrule 
Planet radius & $R_J$ & 1.38 & 1.9 \\
Star radius & $R_{S}$ & 1.118 & 1.657 \\
Planet mass & $M_J$ & 0.69 & 1.47 \\
Semi major axis & AU & 0.047 & 0.0234 \\
Internal temperature & K & 100 & 100 \\
\end{tabular}
\label{tab:params}
\end{table}

All parameters used in the model (\cref{tab:params}) come from The Extrasolar Planet Encyclopaedia\footnote{http://exoplanet.eu/}. For the stellar spectra we use the Kurucz spectra\footnote{http://kurucz.harvard.edu/stars.html} for HD 209458b and the BT-Settl models\footnote{https://phoenix.ens-lyon.fr/Grids/BT-Settl/AGSS2009/} \citep{Allard2012} for WASP-12b. An example of a full steady-state $P$ -- $T$ structure and the effects of pseudo model rotation, using parameters for HD 209458b, can be seen in \cref{fig:fig1}. An atmospheric hotspot is found in the day-side atmosphere, with a temperature inversion corresponding to a \ch{TiO}/\ch{VO} absorption region. The temperature inversion shifts to much lower pressures on the night-side. Of the four positional profiles shown in \cref{fig:fig1}a, the substellar point (0$^\circ$) $P$ -- $T$ profile is closest to the $P$ -- $T$ profile calculated using the 1D time-dependent model. The temperature structure for HD~209458b obtained by our pseudo-2D model is broadly similar to that predicted by 3D GCMs \citep[][see their Figure 3]{Parmentier2013}, for the equatorial region.

It should be reiterated that this simple parameterisation of horizontal advection follows flow in the Lagrangian frame by rotating the column, but does not simulate any material or heat transport horizontally out of the column bounds. It only represents bulk advection by modulating the incident stellar irradiation on the atmospheric column. Hence this model is pseudo-2D, not a full 2D treatment of the atmosphere. In HJ atmospheres, circulation is dominated by the equatorial zonal jet, so pseudo-2D models are useful for these systems.
%
%%%%%%%%%%%%%%%%%%%%%%%%%%%%%%%%%%%%%%%%%%%%%%%%%%%%%%%%%%%%%%%
\subsection{Thermal dissociation and recombination of hydrogen}
\label{subsec:hdis}
\begin{figure*}
\centering
\includegraphics[width=\textwidth]{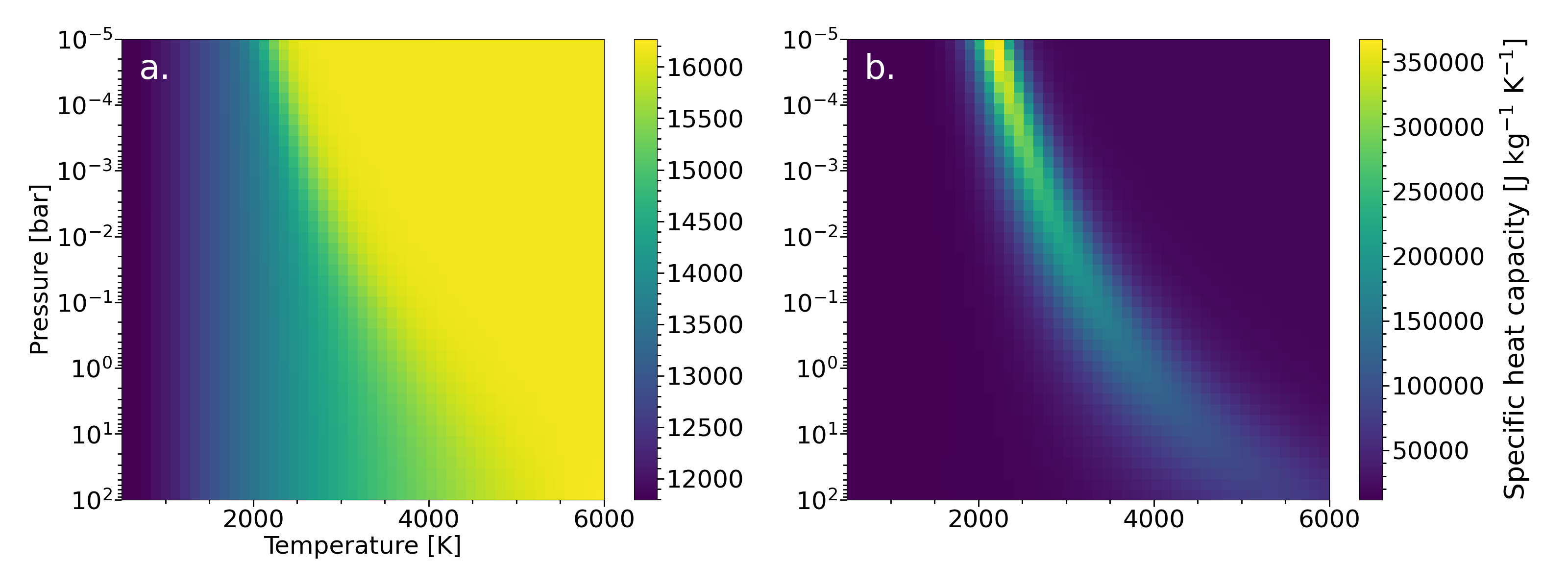}
\caption{Specific heat capacity as a function of pressure and temperature: (a) with only the frozen heat capacity (first term on right of \cref{eq:5}) included and (b) with both the frozen and reaction heat capacities included. Note that the figures have drastically different colour scales, so that variation in the frozen heat capacity can be seen.}
\label{fig:fig2}
\end{figure*}
When molecular hydrogen, \ch{H2}, reaches temperatures between $\sim$2000 -- 4000 K, depending on pressure, it thermally dissociates into atomic hydrogen, H. Atomic hydrogen can then recombine to reform \ch{H2} when the temperature drops, as the reaction is reversible.
\begin{center}
\ch{H_2} (+ M) \ch{ <=> H + H} (+ M).
\end{center}
\noindent
Note that a collision partner (indicated as "M" above) is not needed to calculate the thermodynamics of the thermal dissociation/recombination of hydrogen, but is needed to calculate its kinetics (e.g., the chemical timescales). 

Reaching the required pressure-dependent temperatures for \ch{H2} dissociation is not uncommon in the day-side of UHJ atmospheres. However, the night-side temperature is far colder. As a result, \ch{H2} in the day-side atmosphere thermally dissociates before being pushed onto the night-side by the equatorial jet, where it recombines \citep{2018ApJ...857L..20B}. Due to the endothermic/exothermic nature of these processes, the reactions remove energy from the day-side atmosphere and redistribute it to the night-side. If this energy source/sink is large enough, it affects the temperature structure of the atmosphere. As measurements of transit photometry for UHJs also rely on their relative day-night temperature contrast, any significant changes to the temperature structure from this effect can impact observation.

Hydrogen is by far the most abundant element in UHJ atmospheres, so this dissociation/recombination effect has a large impact on the mean molecular weight within the system, and therefore on its heat capacity. In our pseudo-2D model, we use this to introduce the reaction heat redistribution by adding an additional term in the specific heat capacity \citep{GordonMcBride1994},
\begin{ceqn}
\begin{equation}
c_P = \sum_j n_j C_{P,j}^0 + \sum_j n_j\frac{H_j^0}{T} \Bigg( \frac{\partial\ln{n_j}}{\partial\ln{T}} \Bigg)_P,
\label{eq:5}
\end{equation}
\end{ceqn}
\noindent
where, for a species $j$: $n_j$ is the number of moles per kilogram, $H_j^0$ is the molar enthalpy, and $C_{p,j}^0$ is the molar standard heat capacity.

The first term on the right of \cref{eq:5} is the `frozen' heat capacity, which is simply an abundance weighted sum of the individual heat capacities of the mixture. The second term on the right is the `reaction' heat capacity, which accounts for the temperature-dependent chemistry. This reaction term becomes significant when a species with high molar enthalpy changes abundance over a small temperature range, which is the case for thermal dissociation/recombination of hydrogen. For a full derivation of this equation see \cref{appendix:heat_capacity}. The partial derivative $\left(\frac{\partial\ln{n_j}}{\partial\ln{T}}\right)_P$ is evaluated using a Newton-Raphson iterative method analogous to that used to calculate the equilibrium abundances, as described in \citet[][see their Section 2.5.1]{GordonMcBride1994}.

The specific heat capacity as a function of pressure and temperature (for a solar elemental composition) is shown in \cref{fig:fig2}, when only the frozen heat capacity is included (\cref{fig:fig2}(a)) and when both the frozen and reaction heat capacities are included (\cref{fig:fig2}(b)). The specific heat capacity significantly increases for particular pressure-temperature combinations, corresponding to the dissociation/recombination of \ch{H2}/\ch{H}. This means the reaction only has a significant impact on atmospheric structure when these pressure-dependent temperature combinations are achieved, and the atmosphere horizontally transitions between an \ch{H2} and \ch{H}-dominated regime.

This method of introducing heat redistribution from the dissociation/recombination reaction is equivalent to the method used in \citet{2018ApJ...857L..20B}, whereby the heat redistribution is input directly through the governing energy equation. The tracer method used in \citet{2019ApJ...886...26T} is also equivalent. Our approach is simply the `chemical' method, as opposed to statistical physics. However, by altering the heat capacity calculation, our model is more generalised and can be used to recreate the heat of any other chemical transition reactions we may want to study in the future. We choose to focus on \ch{H2} dissociation/recombination here as it is the most impactful for the $P$ -- $T$ range of UHJs. This method has also been used in the latest version of the Met Office Unified Model (UM), creating a field to implement variable heat capacity \citep{2018A&A...612A.105D}.
%
%%%%%%%%%%%%%%%%%%%%%%%%%%%%%%%%%%%%%%%%%%%%%%%%%%%%%%%%%%%%%%%
\section{The Reaction Heat Redistribution In UHJs}
\label{sec:tdh}
\begin{figure}
\includegraphics[width=0.5\textwidth]{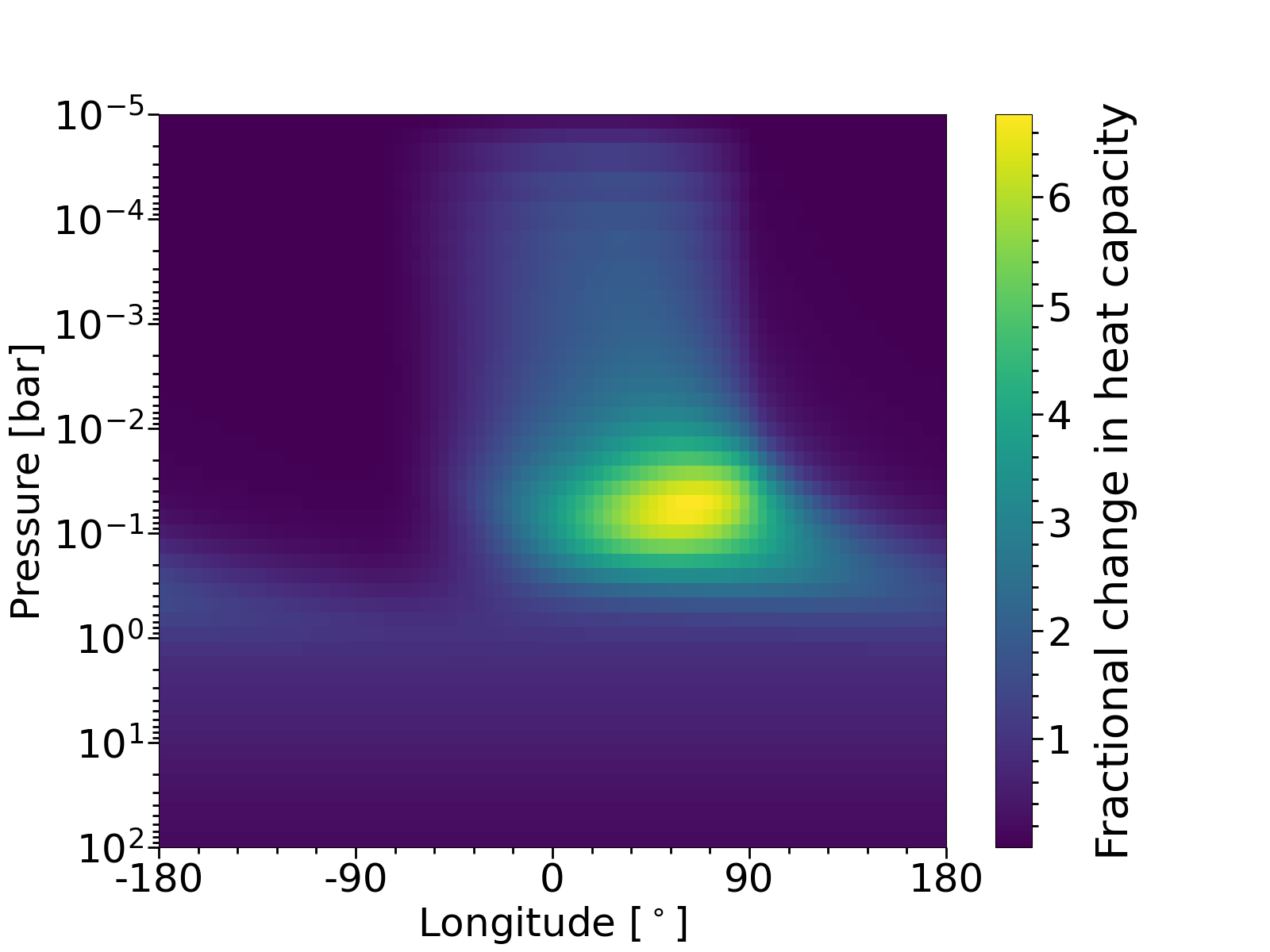}
\caption{Fractional change in the specific heat capacity when including the reaction heat capacity, determined from our pseudo-2D model at steady-state. Wind speed is set at $\sim$5 km s$^{-1}$. \ch{TiO}/\ch{VO} are not included as atmospheric opacity sources.}
\label{fig:fig3}
\end{figure}
\begin{figure*}
\centering 
\includegraphics[width=\textwidth, keepaspectratio]{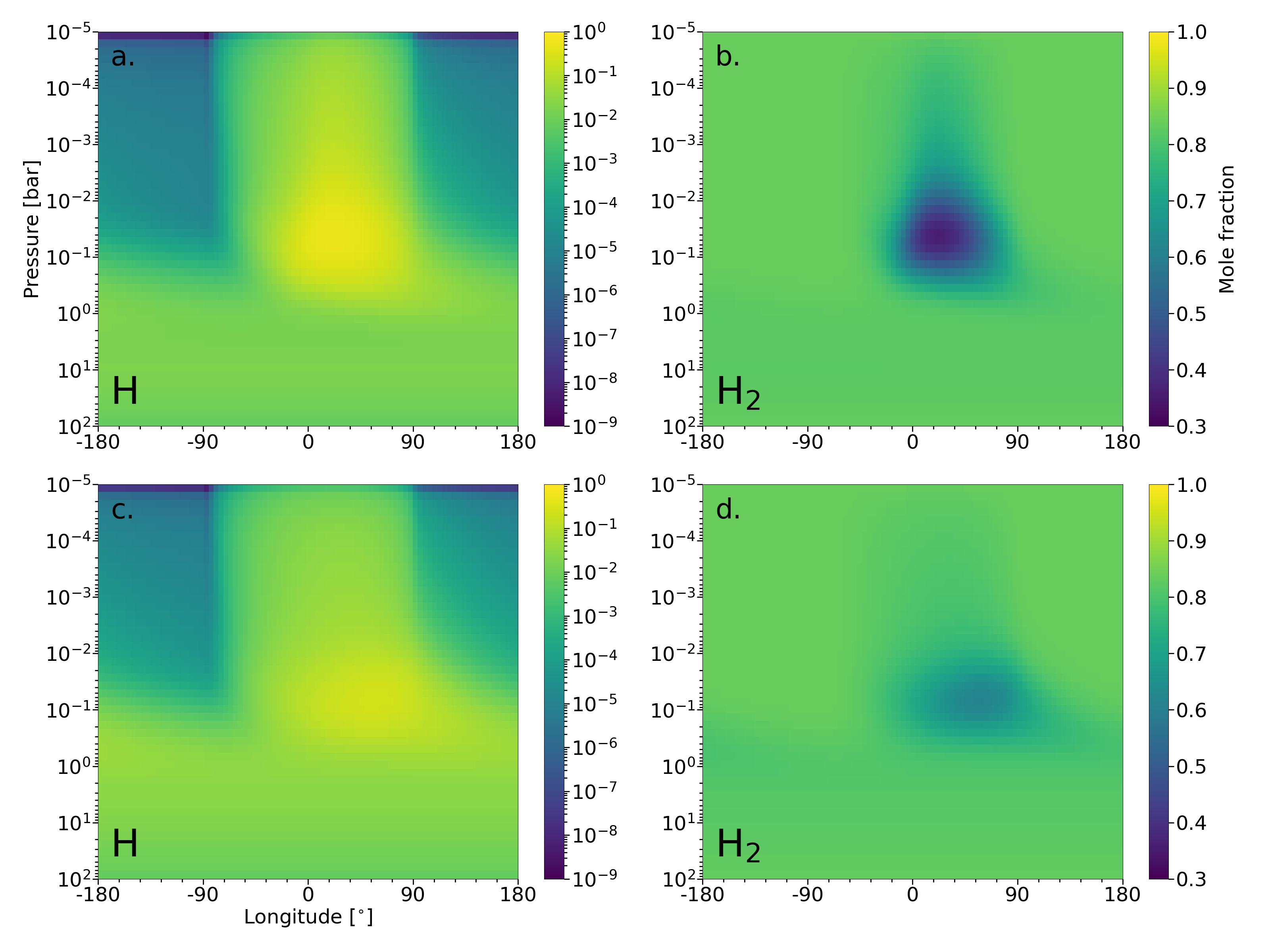}
\caption{Steady-state chemical structure determined from our model: \ch{H} molar fraction (left) and \ch{H2} molar fraction (right), when the reaction heat capacity is excluded (top) and included (bottom), respectively. \ch{TiO}/\ch{VO} are not included as atmospheric opacity sources.}
\label{fig:fig4}
\end{figure*}
\begin{figure*}
\centering 
\includegraphics[width=\textwidth, keepaspectratio]{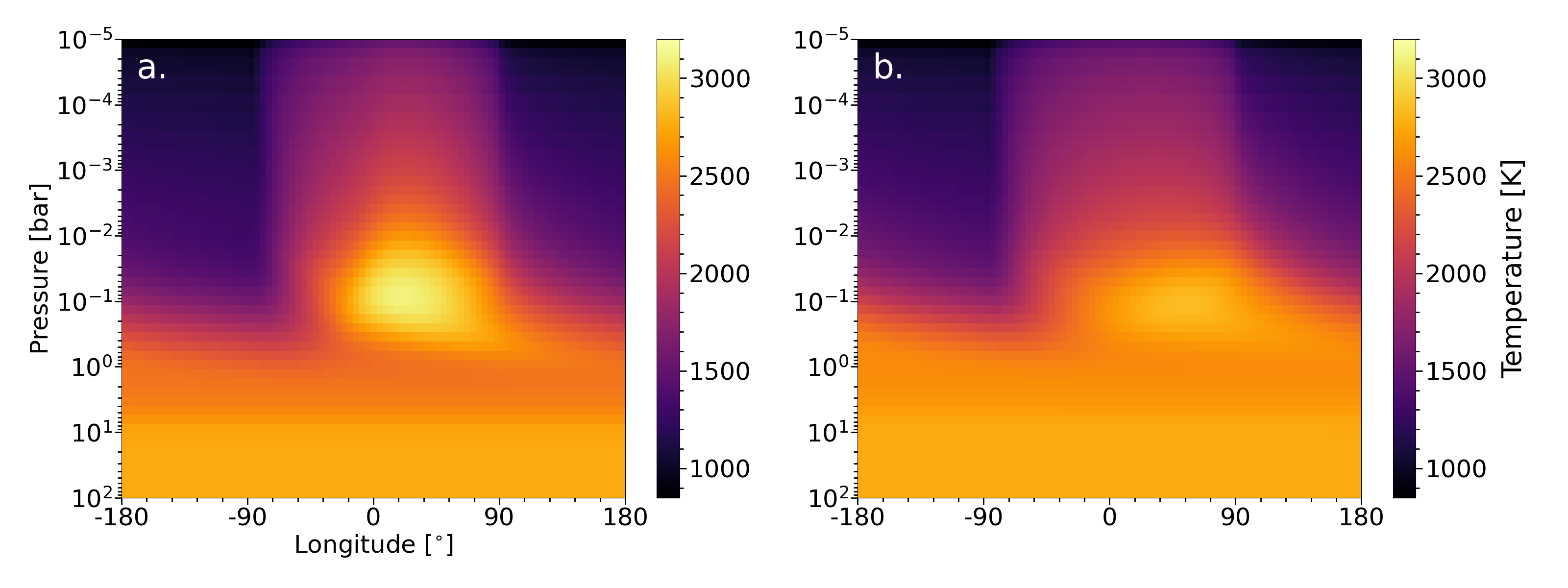}
\caption{Steady-state $P$ -- $T$ structure determined from our model: (a) with the reaction heat capacity excluded and (b) with the reaction heat capacity included. \ch{TiO}/\ch{VO} are not included as atmospheric opacity sources.}
\label{fig:fig5}
\end{figure*}
WASP-12b has an unusually high day-side temperature, in places reaching $\geq 3000$ K \citep{Swain_2013}, and an equilibrium temperature of $\sim$2500 K \citep{2019AJ....158...39C}. This atmosphere represents an interesting case between planetary and stellar conditions. We select WASP-12b as indicative of UHJs, but recognise that in some respects this planet is not typical for this class of objects \citep{2019MNRAS.489.1995B}. However, as we are interested in the mechanism of high temperature thermal dissociation and recombination, and not specifically characterising this planet in detail, omission of the additional processes relevant to WASP-12b will not hinder our wider conclusions.

In our pseudo-2D model, \ch{TiO} and \ch{VO} are initially excluded. The impact of including these species is explored in \cref{subsec:tiovo}. Here we use a timestep of 200 s and a column rotation period of $1.728\times10^{5}$ s, giving a wind speed of $\sim$5 km s$^{-1}$ (4830 m s$^{-1}$), which is scaled proportionally for alternate wind speeds investigated in \cref{subsec:ws}. This is a reasonable estimate given that the zonal-mean zonal wind speed is found to saturate at this value for temperatures above 1500 K \citep{2017ApJ...835..198K}.
%
%%%%%%%%%%%%%%%%%%%%%%%%%%%%%%%%%%%%%%%%%%%%%%%%%%%%%%%%%%%%%%%
\subsection{Excluding TiO/VO}
\begin{figure}
\centering
\includegraphics[width=0.5\textwidth]{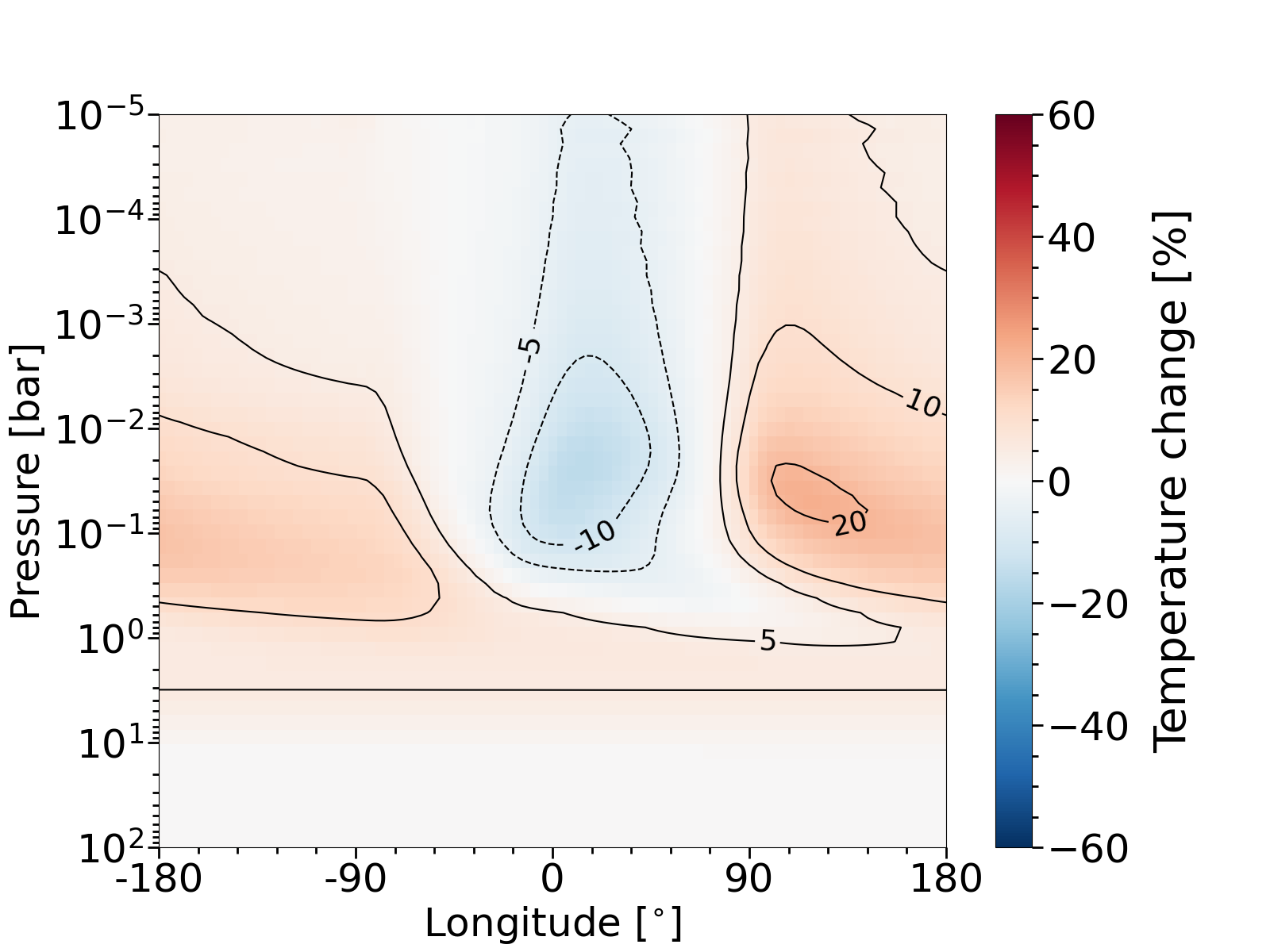}
\caption{Percentage change in temperature when including the reaction heat capacity, corresponding to the difference between \cref{fig:fig5}(a) and \cref{fig:fig5}(b). \ch{TiO}/\ch{VO} are not included as atmospheric opacity sources.}
\label{fig:fig6}
\end{figure}
\begin{figure}
\includegraphics[width=0.5\textwidth]{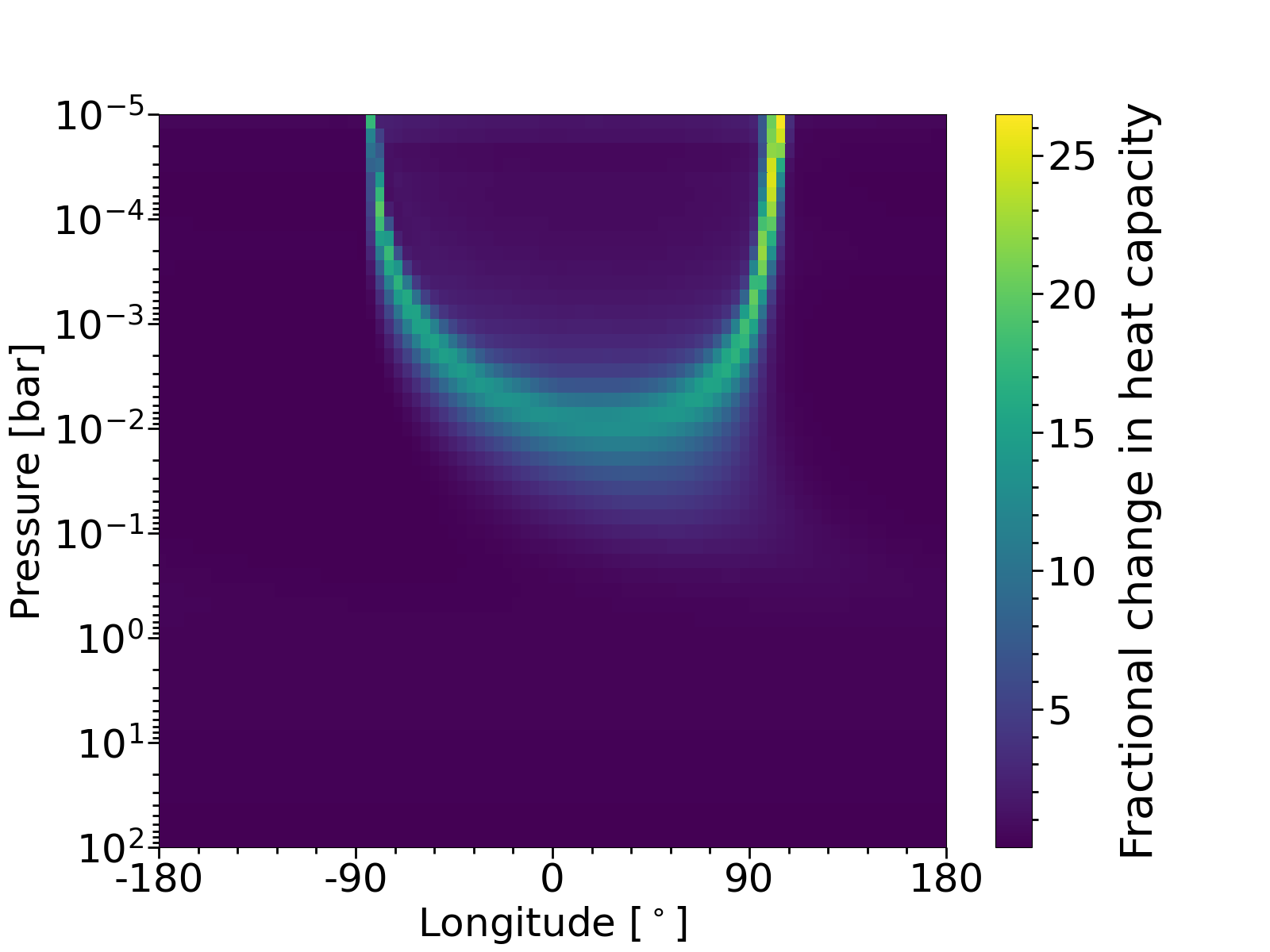}
\caption{Fractional change in the specific heat capacity when including the reaction heat capacity, determined from our pseudo-2D model at steady-state. Wind speed is set at $\sim$5 km s$^{-1}$ and \ch{TiO}/\ch{VO} are included as atmospheric opacity sources.}
\label{fig:fig7}
\end{figure}
\begin{figure*}
\centering 
\includegraphics[width=\textwidth, keepaspectratio]{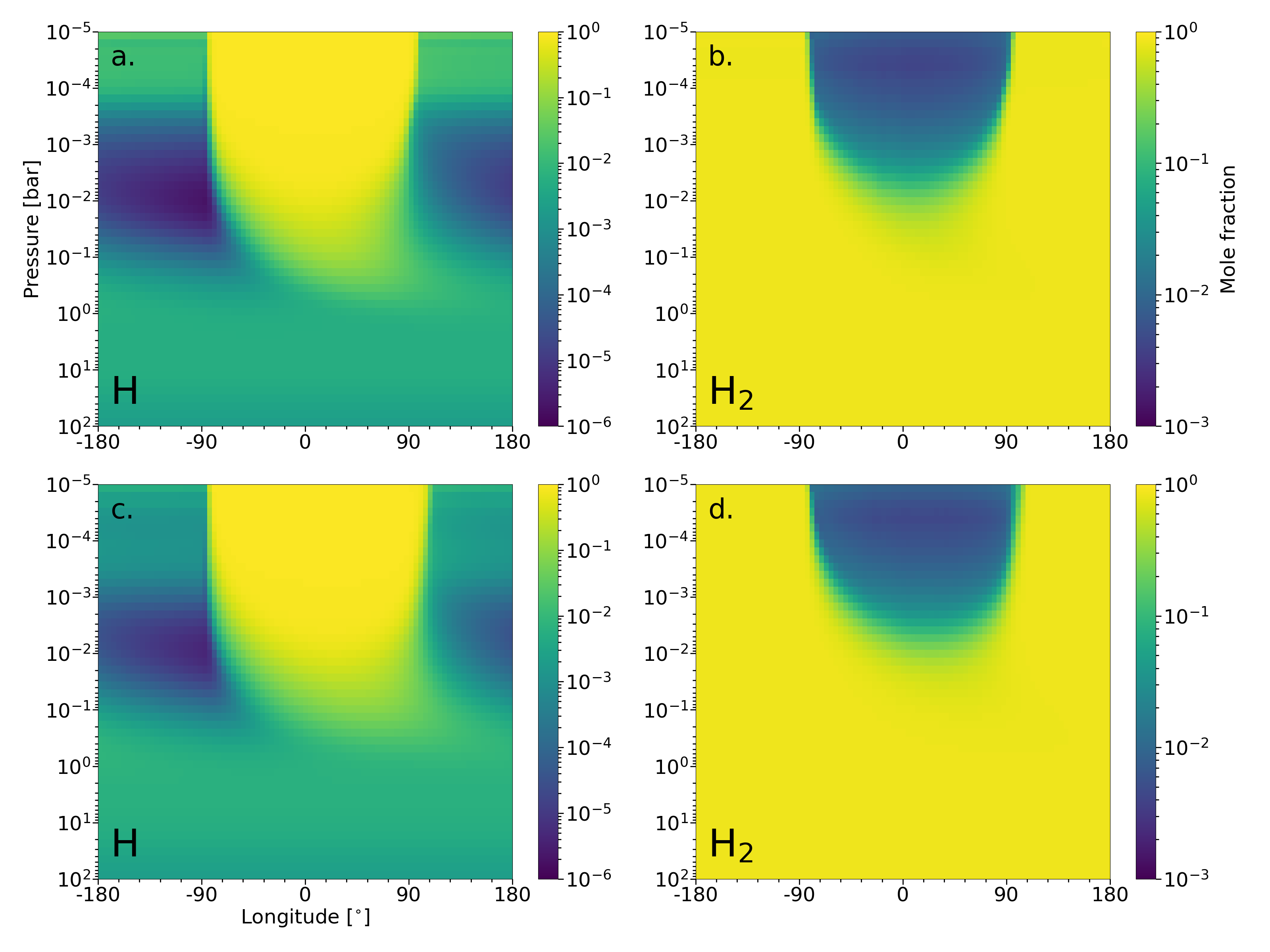}
\caption{Steady-state chemical abundances determined from our model: \ch{H} molar fraction (right) and \ch{H2} molar fraction (left), when the reaction heat capacity is excluded (top) and included (bottom), respectively. \ch{TiO}/\ch{VO} are included as atmospheric opacity sources.}
\label{fig:fig8}
\end{figure*}
\begin{figure*}
\centering 
\includegraphics[width=\textwidth, keepaspectratio]{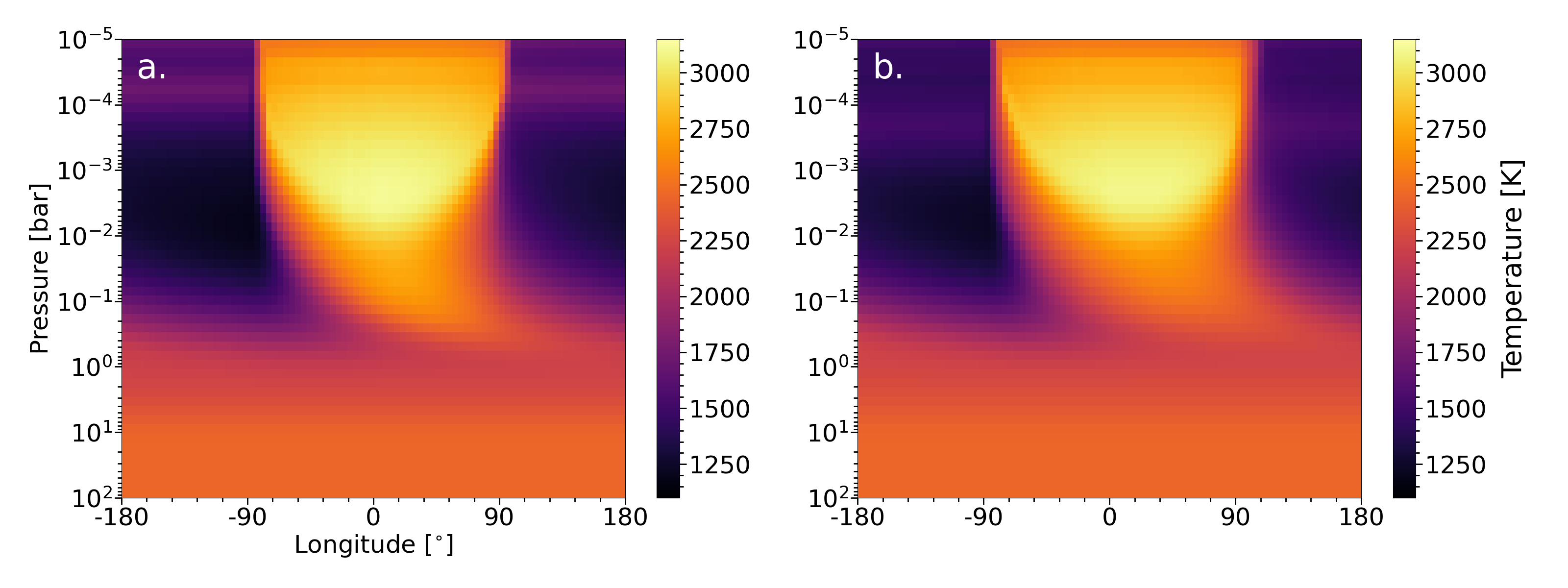} 
\caption{Steady-state $P$ -- $T$ structure determined from our model: (a) with the reaction heat capacity excluded and (b) with the reaction heat capacity included. \ch{TiO}/\ch{VO} are included as atmospheric opacity sources.}
\label{fig:fig9}
\end{figure*}
To analyse the effects of the reaction heat redistribution within this atmosphere, the pseudo-2D model is first run without the reaction heat capacity included (\cref{eq:5}). The initial profile is calculated from the 1D time-dependent model (\cref{subsec:tdmodel}). The model is then run again, this time including the reaction heat capacity. \cref{fig:fig3} shows the fractional change in steady-state heat capacity structure once the reaction heat capacity is applied. The fractional change in steady-state chemical abundances of \ch{H}/\ch{H2} and corresponding $P$ -- $T$ structure can be seen in \cref{fig:fig4} and \cref{fig:fig5} respectively.

First, as shown in \cref{fig:fig3}, we see that when the reaction heat capacity is applied the atmospheric heat capacity increases drastically on the day-side, most prominently between pressures of $10^{-1}$ -- $10^{-2}$ bar. This region corresponds to an area in the atmosphere at which the largest variations in \ch{H2} and \ch{H} abundance occur, due to the dissociation/recombination affect, and as a result the $P$ -- $T$ profile aligns with the reaction heat capacity's optimal $P$ -- $T$ range (seen in \cref{fig:fig2}).

This can clearly be seen when inspecting the steady-state chemical distributions found in \cref{fig:fig4}. As thermal dissociation of \ch{H2} occurs, between the substellar point and evening terminator, it cools the day-side atmosphere. Consequently the temperature is lowered and falls below the optimal reaction $P$ -- $T$ range, thus reducing the amount of dissociation occurring. This means, when including the reaction heat capacity, there is an increased abundance of \ch{H2} at the peak dissociation point, with a decrease at the peak recombination point and throughout the night-side atmosphere. The same logic, in reverse, applies to the H abundance. Recombination heating leads to increased abundance just beyond the evening terminator and throughout the night-side, but a decrease in the day-side atmosphere between the substellar point and evening terminator.

To see the effects of heat redistribution caused by the reaction heat capacity, we can look at a comparison between the steady-state $P$ -- $T$ structures (\cref{fig:fig5}). Firstly, it is interesting to note that a slight temperature inversion is present in the atmosphere at $10^{-1}$ bar, despite the lack of \ch{TiO} and \ch{VO} as opacity sources. This temperature inversion occurs as the radiative timescale above the $\sim$ 1 bar pressure level is short enough for the day-side atmosphere to respond to stellar heating. At higher pressures, the radiative timescale is much longer than the rotation timescale so the atmosphere does not respond rapidly enough to the radiative heating, and a temperature inversion forms. This inversion is decreased when the reaction heat capacity is included in our model. As the radiative timescale is roughly proportional to the atmospheric heat capacity, increased heat capacity due to chemical transition reactions makes the atmosphere more stable against the radiative heating. In other words, a higher incident energy would be required to maintain the temperature inversion for the same wind speed.

By calculating the relative change in longitudinal temperature, seen in \cref{fig:fig6}, we observe a clear reduction in the global temperature gradient, as the day-side cools and night-side warms. This is, unsurprisingly, most prominent between $10^{-1}$ -- $10^{-2}$ bars, where we observe the largest horizontal chemical variation. In this case, the atmosphere between the substellar point and evening terminator is seen to cool by over $\sim$10\% in places, or $\sim$400 K. The night-side atmosphere is even more drastically affected, heating by up to $\sim$20\% just beyond the evening terminator. This also corresponds to a temperature change of $\sim$400K, as the percentage change is relative to the original temperature structure. Perhaps the most interesting consequence of these changes in temperature is a significant overall shift in the atmospheric hotspot towards the evening terminator.

In our model, the temperature changes are mostly confined to the upper regions of the atmosphere, at pressures below 1 bar, where strong irradiation can penetrate. It should again be mentioned that our model lacks time-dependent non-equilibrium reactions, as we assume local chemical equilibrium, meaning thermal dissociation/recombination occurs instantly. If a more realistic timescale for the thermal dissociation/recombination were to be considered, the atmospheric hotspot would likely shift even further from the substellar point, depending on the competition between chemical, radiative and advective timescales.
%
%%%%%%%%%%%%%%%%%%%%%%%%%%%%%%%%%%%%%%%%%%%%%%%%%%%%%%%%%%%%%%%
\subsection{Effects of \ch{TiO} and \ch{VO}}
\label{subsec:tiovo}
\begin{figure}
\centering 
\includegraphics[width=0.5\textwidth]{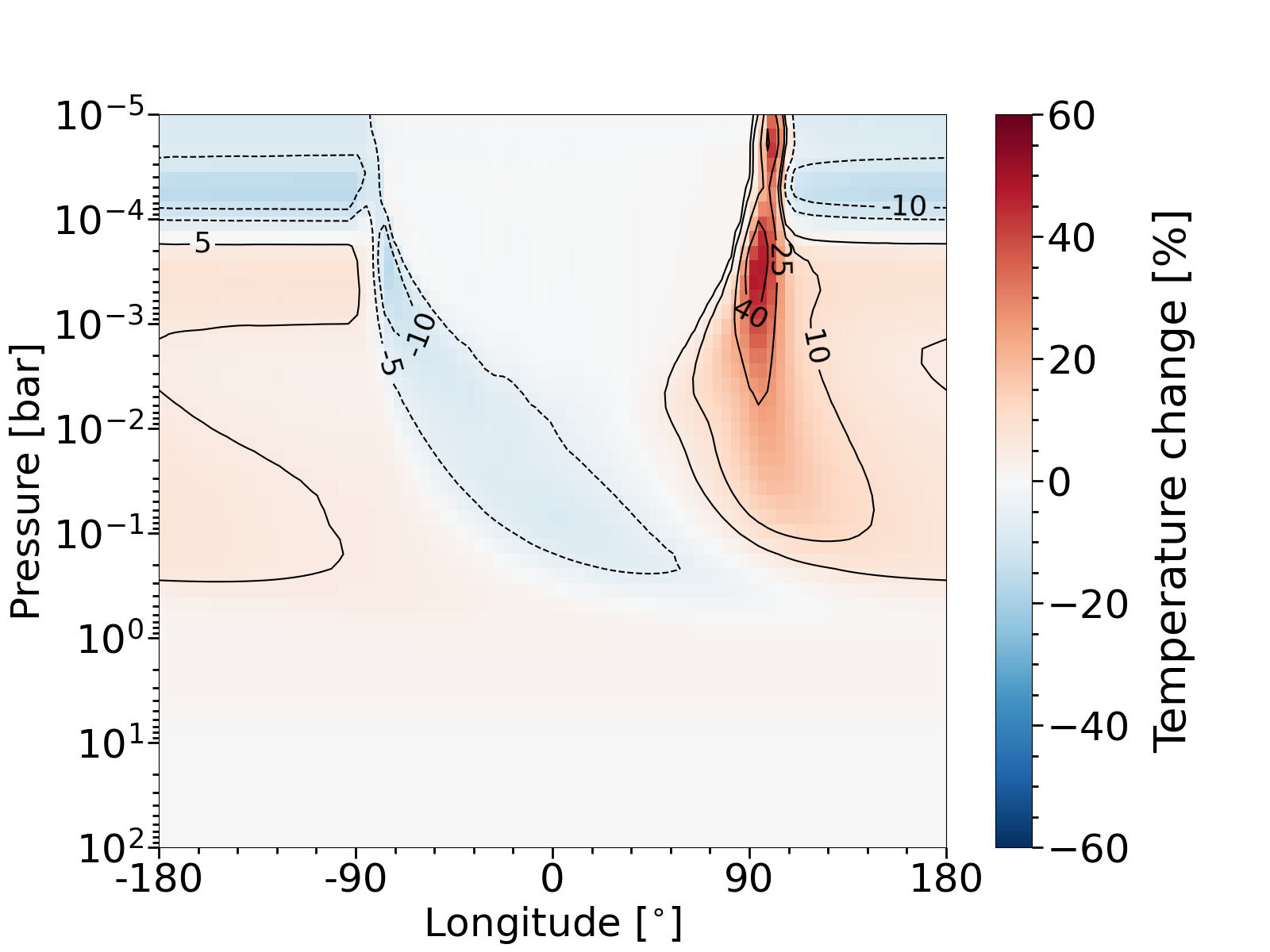}
\caption{Percentage change in temperature when including the reaction heat capacity, corresponding to the difference between \cref{fig:fig9}(a) and \cref{fig:fig9}(b). \ch{TiO}/\ch{VO} are included as atmospheric opacity sources.}
\label{fig:fig10}
\end{figure}
\ch{TiO} and \ch{VO} are thought to be the primary drivers for steep temperature inversions observed on the day-side of HJ atmospheres \citep{2008A&A...492..585D}. Both \ch{TiO} and \ch{VO} display strong photo-absorption in the near UV/optical range, causing strong heating in the upper atmosphere, and acting as a soft irradiation barrier to lower altitudes. However, there is an ongoing debate about the role of \ch{TiO}/\ch{VO} in atmospheres with particularly large day-night temperature contrasts. Although previous studies have shown that these compounds are likely to be thermally dissociated on the hotter day-side, specifically for WASP-121b which has a similar (albeit slightly cooler) $P$ -- $T$ structure than WASP-12b \citep{2018A&A...617A.110P}, they may condense out of the night-side atmosphere entirely \citep{{2018A&A...617A.110P,Merritt_2020}}. Furthermore, both \ch{TiO} and \ch{VO} are currently very difficult to detect directly \citep{Mikal-Evans2020}, as this would require high resolution observations in the near-UV/optical range. In this section, we investigate the effects of the reaction heat redistribution on atmospheric structure, with \ch{TiO} and \ch{VO} included in our model.

The fractional change in steady-state heat capacity structure, now including \ch{TiO} and \ch{VO}, can be seen in \cref{fig:fig7}. From this we see that the heat capacity now drastically increases in a thin band within the day-side atmosphere at pressures lower than $10^{-2}$ bar. This corresponds to the edge of the \ch{TiO}/\ch{VO} photo-absorption region. Because of the temperature inversion, the pressure-dependent temperature for the reaction heat capacity is now met predominantly at the peak of \ch{TiO}/\ch{VO} absorption. This can clearly be seen in the steady-state distributions of both \ch{H} and \ch{H2} abundance (\cref{fig:fig8}), where the vast majority of thermal \ch{H2} dissociation is now confined to the day-side atmosphere above the $10^{-2}$ bar pressure level. However, the dissociation/recombination has an increasingly limited impact on the heat capacity inside the region, although the optimal pressure-dependent temperature is met throughout, as the \ch{TiO}/\ch{VO} opacities almost entirely dominate the temperature structure. The result is a large increase to the heat capacity in a band around the edges of the \ch{TiO}/\ch{VO} absorption region; with changes on the morning edge caused by \ch{H2} dissociation and on the evening edge caused by \ch{H} recombination. Some \ch{H2} dissociation is also found to occur in the upper atmosphere, above the $10^{-4}$ bar pressure level, as strong absorption by \ch{TiO}/\ch{VO} maintains a small temperature inversion throughout the night-side atmosphere and the temperature required for the \ch{H2} to \ch{H} transition reduces towards lower pressures. This causes some cooling at these altitudes on the night-side.

By looking at the the periodic steady-state temperature structures for the simulations (\cref{fig:fig9}), we can see the \ch{H2} dissociation therefore reduces the magnitude of the temperature inversion on the morning edge of the hotspot and shifts its peak to lower-pressure levels. Conversely, on the evening side of the hotspot, the thermal H recombination heating contributes to significantly increase the temperature at the edge of the \ch{TiO}/\ch{VO} absorption region. As previously seen in the case when \ch{TiO}/\ch{VO} have been excluded, this again causes a strong shift in the atmospheric hotspot towards the evening terminator.

The dominance of \ch{TiO}/\ch{VO} within the primary dissociation region also results in a significant disparity between the heating and cooling in this case. From \cref{fig:fig10} we can see there is a large magnitude increase in the heating peak, from $\sim$20\% ($\sim$400 K) up to over $\sim$40\% ($\sim$800 K), when comparing with the \ch{TiO}/\ch{VO} exclusion case. Whereas the cooling remains at a similar value of $\sim$5 -- 10\% (between $\sim$200 -- 400 K). We do, however, now see equal cooling at high altitudes in the night-side atmosphere, as predicted by the increased \ch{H} abundance seen in \cref{fig:fig8}.
%
%%%%%%%%%%%%%%%%%%%%%%%%%%%%%%%%%%%%%%%%%%%%%%%%%%%%%%%%%%%%%%%
\subsection{Effects of wind speed}
\label{subsec:ws}
\begin{figure*}
\centering 
\includegraphics[width=\textwidth]{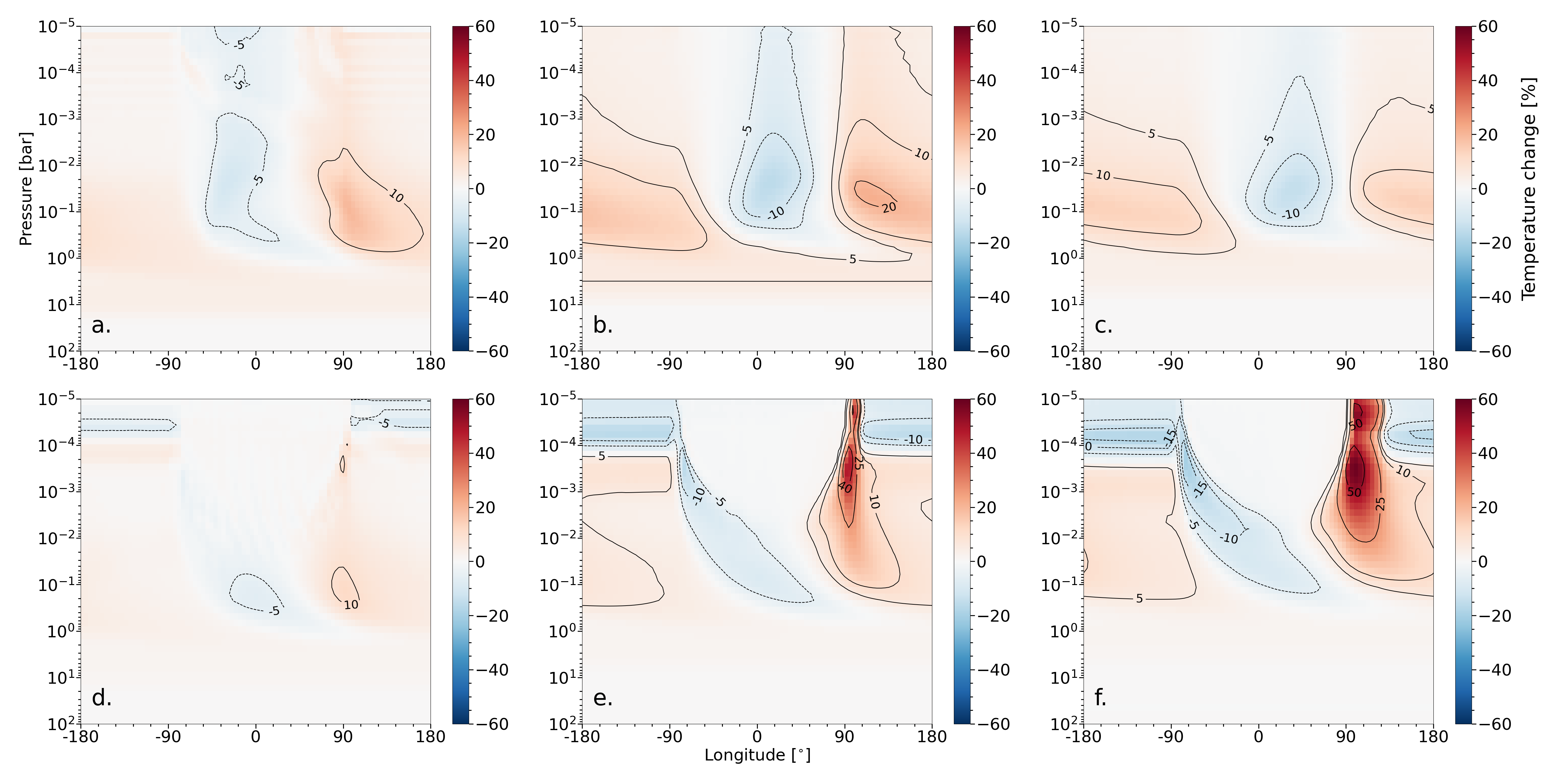}
\caption{Percentage change in temperature when including the reaction heat capacity with \ch{TiO}/\ch{VO} either excluded (top) or included (bottom) as atmospheric opacity sources. Wind speeds were set at $\sim$1, $\sim$5 and $\sim$10 km s$^{-1}$, from left to right respectively.}
\label{fig:fig11}
\end{figure*}
In order to consider the sensitivity of the reaction heat redistribution to our selected wind speed, the model is re-run a number of times with different column rotation periods, to simulate different dynamical conditions excluding, and then including, the reaction heat capacity. For each column rotation period, the resulting percentage temperature change has then been determined (\cref{fig:fig11}). This is done both excluding and including \ch{TiO}/\ch{VO}.

As wind speed is increased from $\sim$1 km s$^{-1}$ up to $\sim$10 km s$^{-1}$ (R=864000~s and R=86400~s are used, giving speeds of 966 m s$^{-1}$ and 9660 m s$^{-1}$ respectively), some key trends become apparent. Other than for the 10 km s$^{-1}$ \ch{TiO}/\ch{VO} exclusion case, which is discussed shortly, the magnitude of both relative cooling and heating from the reaction heat redistribution is seen to increase with increasing wind speed in both cases excluding and including \ch{TiO}/\ch{VO}. Heating from \ch{H} recombination is also consistently affected more than the cooling from \ch{H2} dissociation in the presence of \ch{TiO}/\ch{VO}, for the same reason as discussed in Section \ref{subsec:tiovo} (dominance of \ch{TiO}/\ch{VO} opacity in the primary photo absorption region). Protracted heating, further from the primary recombination zone, and low pressure cooling ($\leq10^{-4}$ bar) throughout the night-side atmosphere is additionally increased with higher wind speeds.

These effects can be explained by considering the balance between advective and radiative timescales within the atmosphere. As the wind speed is set by the rotation rate of the column, this rate emulates the advective timescale which decreases as the model wind speed is increased. The atmospheric temperature takes a finite amount of time to respond to heating, and the radiative timescale is independent of rotation, so the radiative heating has diminishing effects at higher wind speeds as the advective timescale becomes comparatively small. The chemical timescale, however, is time-independent regardless of the model wind speed, due to the assumption of chemical equilibrium. This means that, at higher wind speeds, the cooling and heating from the reaction heat redistribution increasingly outweigh the radiative heating and cooling within the atmosphere, and therefore have a progressively larger effect on the $P$ -- $T$ structure. 

The exception to this trend is seen when comparing the increase from 5 to 10 km s$^{-1}$ in the \ch{TiO}/\ch{VO} exclusion case. Here the reaction heat redistribution is actually observed to decrease at the higher wind speed. This can be explained by considering the effects that a faster jet has on the atmosphere. The temperature inversion decreases as the dynamical timescale begins to outweigh the radiative timescale and the day-night temperature contrast decreases as the faster jet more thoroughly homogenises the atmosphere. As a consequence, the decreased temperature progressively falls below the optimal temperatures for the \ch{H_2}/\ch{H} thermal transition, effectively capping the associated reaction heat capacity at a value set by the threshold wind speed and minimising the overall impact of the reaction heat redistribution. This threshold effect is not observed when \ch{TiO}/\ch{VO} are present, as in this case the temperature inversion is set by the photo-absorption of these species rather than purely by radiative heating, so these chemicals stabilise the day-side atmosphere at the required transition temperature.

Unfortunately, the pseudo-2D radiative solver currently lacks any dynamical or chemical feedback on the wind speed, so it is unclear if the reaction heat redistribution would directly impact the jet, particularly at the equator where it is strongest. Conversely, the wind speed could be the primary moderator within the system, and control the dissociation/recombination fractions. Studies of this effect using a 3D GCM have shown the speed of the equatorial jet to decrease if the rotation period is fixed \citep{2019ApJ...886...26T}. However, we show here that chemical feedback is also very important. To fully explore the dynamics, a 3D global circulation model, with time-dependent chemical feedback, would need to be used. This is a promising avenue to explore in the future, but is not possible until the treatment of chemical feedback within higher dimensionality models is much improved.
%
%%%%%%%%%%%%%%%%%%%%%%%%%%%%%%%%%%%%%%%%%%%%%%%%%%%%%%%%%%%%%%%
\subsection{Consequences for the pseudo-2D phase curve}
\label{subsec:pc}
\begin{figure*}
\centering 
\includegraphics[width=\textwidth]{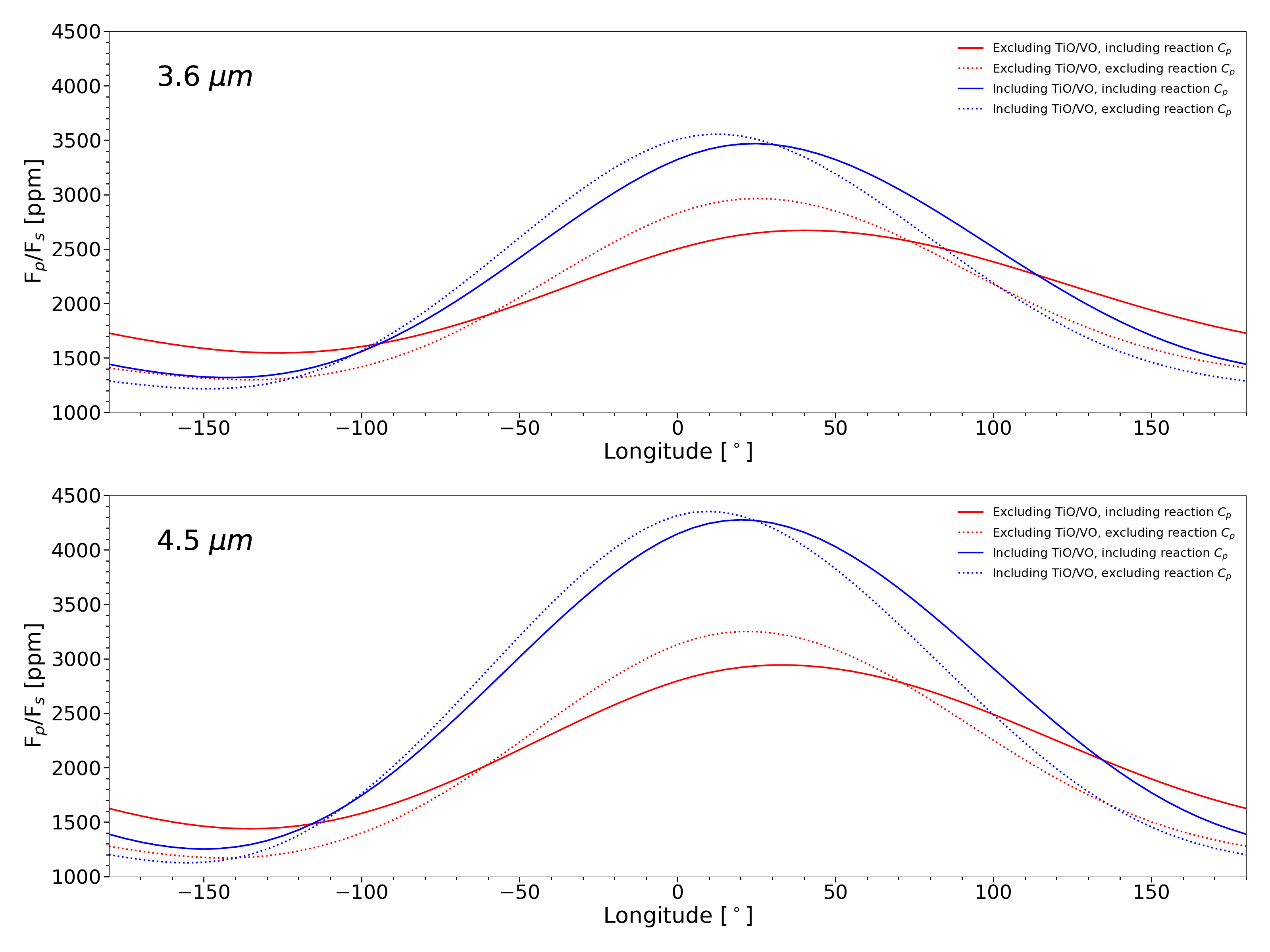}
\caption{Pseudo-2D phase curves determined from the emission spectra calculated by our model (see \cref{appendix:spectra}), at 3.6~$\mu$m and 4.5~$\mu$m, with the reaction heat capacity either excluded (dashed lines) or included (solid lines) and with TiO/VO either excluded (red lines) or included (blue lines) as atmospheric opacity sources. Outputs from the 5 km s$^{-1}$ wind speed models were used.}
\label{fig:fig12}
\end{figure*}
By computing the emission spectra for individual steady-state $P$ -- $T$ profiles in the atmosphere at $5^{\circ}$ intervals, using 500-band correlated-$k$ cross-sections, we produce pseudo-2D phase curves for our model. A sample of these emission spectra can be seen in \cref{appendix:spectra}. These pseudo-2D phase curves, at 3.6~$\mu$m and 4.5~$\mu$m for cases both excluding and including the \ch{TiO}/\ch{VO} species and reaction heat capacity, are shown in \cref{fig:fig12}. Given our model assumptions (i.e equilibrium chemistry, zero convection, no dynamical feedback, no clouds or haze) and its 2D nature, meaning only longitudinal contributions to the atmospheric emission are considered, these cannot be considered as true phase curves and should not be taken as completely realistic. They can, however, provide a very reasonable approximation to how \ch{H2} dissociation/recombination may impact observation, especially in the presence of \ch{TiO}/\ch{VO}. Phase curve data in \citet{2019MNRAS.489.1995B} shows roughly a 4000 ppm ratio in day-side to stellar flux at 3.6~$\mu$m and 4500 ppm at 4.5~$\mu$m. This aligns well with our model, particularly in the cases including \ch{TiO}/\ch{VO} where we calculate $\sim$3500 ppm and $\sim$4500 ppm ratios respectively. In both scenarios we explore, the spectral flux amplitude damping we would expect to observe due to the reduction in global temperature gradient is present. This damping is much more significant in the case excluding \ch{TiO}/\ch{VO}, $\sim$400 ppm ($\sim$13.5\%) compared to $\sim$100 ppm ($\sim$3\%) at 3.6~$\mu$m and $\sim$300 ppm ($\sim$9\%) compared to $\sim$75 ppm ($\sim$2\%) at 4.5~$\mu$m, which is also to be expected as we have already seen the reaction heat redistribution to have a more protracted impact on the atmosphere in this instance. This damping has been studied in a number of previous papers, and has been used to explain some flattening observed in UHJ phase curves \citep{2019ApJ...886...26T}.

The other notable feature in both pseudo-2D phase curves is a clear phase offset increase when the reaction heat redistribution is considered. This offset increase is very similar in both cases at 3.6~$\mu$m, with a value of $\sim$15$^{\circ}$. At 4.5~$\mu$m it is slightly more significant in the \ch{TiO}/\ch{VO} exclusion case, where the offset is increased by $\sim$15$^{\circ}$, compared to $\sim10^{\circ}$ when \ch{TiO}/\ch{VO} are included. Of course, this will also vary significantly on a case by case basis, and be heavily influenced under more rigorous dynamical treatment. As we have seen, increasing wind speeds push the hotspot, and the resulting phase offset, further from the substellar point as the reaction heat redistribution increasingly outweighs the radiative temperature changes.
%
%%%%%%%%%%%%%%%%%%%%%%%%%%%%%%%%%%%%%%%%%%%%%%%%%%%%%%%%%%%%%%% 
\section{Conclusions}
\label{sec:conc}
During this study a time-dependent pseudo-2D model for ultra-hot Jupiter exoplanets has been constructed and used, based on the 1D/2D atmosphere model ATMO. This new model is capable of accounting for the heating/cooling effect of any chemical transition reactions, and has been applied to the heat redistribution caused by the thermal dissociation and recombination of hydrogen, under time-independent equilibrium chemical conditions. From the results of this study, the following key conclusions can be drawn:

\begin{itemize}
\item Building upon the work of \citet{2018ApJ...857L..20B}, displaying the importance of \ch{H2} dissociation/recombination in atmospheres exceeding 2000K, we have shown how the presence of certain species (\ch{TiO},\ch{VO}) changes the impact of any resulting heat redistribution.

\item The day-side atmospheric temperature is found to be cooled by up to $\sim$10\% ($\sim$400 K) due to thermal \ch{H2} dissociation in our model. Thermal H recombination heating has an even higher relative impact, reaching over $\sim$20\% ($\sim$400 K) just beyond the evening terminator. With TiO/VO included within the model, cooling at the \ch{TiO}/\ch{VO} absorption peak is seen to remain at a similar value of $\sim$10\% ($\sim$400 K), due to the dominance of \ch{TiO}/\ch{VO} opacity within the photo-absorption region. However, heating at the evening terminator is found to be much higher at over 40\% ($\sim$800 K).

\item The reaction heat redistribution causes a shift in the atmospheric hotspot towards the evening terminator, which is found to be more significant when excluding \ch{TiO}/\ch{VO} from our model. This consequently results in a positive shift in phase offset for the pseudo-2D phase curves. The increased heat redistribution has also been demonstrated to cause significant spectral flux amplitude damping in the pseudo-2D phase curves. This amplitude damping is far more significant when \ch{TiO} and \ch{VO} are not present.

\item With \ch{TiO} and \ch{VO} excluded, increasing the model wind speed, up to a certain threshold value, has been found to increase the magnitude of atmospheric heating and cooling and the distance of protracted heating and cooling throughout the night-side atmosphere. At a threshold wind speed, the reaction heat redistribution is capped by reducing atmospheric temperatures and above this the magnitude of both heating and cooling decreases as the atmosphere becomes increasingly homogenised. \ch{TiO}/\ch{VO} photo-absorption stabilises hotspot temperatures, meaning that in the presence of these species the \ch{H_2}/\ch{H} transition temperatures are achieved regardless of wind speed. Protracted night-side cooling at pressures below $10^{-4}$ bar is also observed in this case. Any increase in the heating and cooling acts to increase the resulting hotspot shift at higher wind speeds (or lower the shift above the threshold wind speed when \ch{TiO}/\ch{VO} are not present). Although it is noted that a more rigorous investigation, one including dynamical feedback, into this effect is required to draw more robust conclusions.
\end{itemize}

\noindent
There are of course limitations to our model which could be addressed. By far the most significant omission comes from inherent limitations due to dimensionality, and therefore limited dynamical feedback, with the wind speed being a prescribed model parameter. Because of this, the relative heating and cooling calculated here is likely to be much larger than expected and observed. Extending our model into 3D would drastically reduce both, as the reaction has its strongest effect at the equator. The presence of night-side clouds, which are thought to significantly impact atmospheric dynamics and radiative transfer in hot Jupiters \citep{2020arXiv201103302H}, are also not included here. These clouds have been shown capable of altering observational phase offset by counterbalancing day-side atmospheric emission, meaning that the offset does not necessarily track the atmospheric hotspot \citep{2020MNRAS.tmp.3310P}.

Any variations in planetary equilibrium temperature, and resulting day-night temperature gradient, would also alter the reaction heat redistribution. The parameters used in our model, roughly simulating WASP-12b, highlight a particularly extreme case, with a much higher equilibrium temperature than average, so we would expect most UHJ exoplanets to have a decreased global temperature gradient and therefore {display diminishing effects from the reaction heat redistribution}. Some assumptions, such as chemical equilibrium, would have only a limited effect on the transition reaction. The inclusion of time-dependent non-equilibrium chemical timescales would act to increase the protracted nature of heating and cooling, as in our model any dissociation/recombination happens instantly, although this would only be significant for high wind speeds due to the fast chemical timescales. For a wind speed of $\sim$5 km s$^{-1}$, the \ch{H2}-\ch{H} interconversion chemical timescales is calculated, on the basis of the present equilibrium abundances, to be shorter than the advection timescale below the $10^{-5}$ bar pressure level at 0$^{\circ}$ longitude (for \ch{H2}$\rightarrow$\ch{H}+\ch{H}) and below the $10^{-2}$ bar pressure level at 180$^{\circ}$ longitude (for \ch{H}+\ch{H}$\rightarrow$\ch{H2}). Variation in planetary mass, and corresponding surface gravity would also have a significant impact, see \cref{appendix:grav} for a brief discussion on this. Interactions between the highly ionised atmosphere and strong magnetic field have also been shown to impact the phase offset in drastic and unpredictable ways \citep{2014ApJ...794..132R}. These magnetic effects are far beyond the scope of this paper, but do provide a promising further avenue for future improvements in UHJ modelling.

Fortunately, very recent progress within Exeter's exoplanet group has provided breakthroughs when running consistent chemistry with 3D models \citep{2018ApJ...855L..31D,2020A&A...636A..68D}. This study could therefore soon be repeated, utilising the Met Office's 3D atmosphere model \citep{2014A&A...561A...1M}, to more thoroughly investigate feedback between the reaction heat recirculation and atmospheric dynamics, whilst maintaining chemical consistency.
\section*{Acknowledgements}
The authors wish to acknowledge the support of the Department of Physics at the University of Exeter, and all the staff involved in the MPhys programme, through which this work was initially undertaken. AR acknowledges support from a Science and Technology Facilities Council Scholarship (20/21\_STFC\_1389376). BD, EH and NJM acknowledge support from a Science and Technology Facilities Council Consolidated Grant (ST/R000395/1). NJM also acknowledges funding through the UKRI Future Leaders Scheme: MR/T040866/1. PT acknowledges support by the European Research Council under grant agreement ATMO 757858.

%%%%%%%%%%%%%%%%%%%%%%%%%%%%%%%%%%%%%%%%%%%%%%%%%%
\section*{Data Availability}

The data underlying this article will be shared on reasonable request to the corresponding author.

%The inclusion of a Data Availability Statement is a requirement for articles published in MNRAS. Data Availability Statements provide a standardised format for readers to understand the availability of data underlying the research results described in the article. The statement may refer to original data generated in the course of the study or to third-party data analysed in the article. The statement should describe and provide means of access, where possible, by linking to the data or providing the required accession numbers for the relevant databases or DOIs.

%%%%%%%%%%%%%%%%%%%% REFERENCES %%%%%%%%%%%%%%%%%%

% The best way to enter references is to use BibTeX:

\bibliographystyle{mnras}
\bibliography{bibliography} % if your bibtex file is called example.bib

\begin{thebibliography}{}
\makeatletter
\relax
\def\mn@urlcharsother{\let\do\@makeother \do\$\do\&\do\#\do\^\do\_\do\%\do\~}
\def\mn@doi{\begingroup\mn@urlcharsother \@ifnextchar [ {\mn@doi@}
  {\mn@doi@[]}}
\def\mn@doi@[#1]#2{\def\@tempa{#1}\ifx\@tempa\@empty \href
  {http://dx.doi.org/#2} {doi:#2}\else \href {http://dx.doi.org/#2} {#1}\fi
  \endgroup}
\def\mn@eprint#1#2{\mn@eprint@#1:#2::\@nil}
\def\mn@eprint@arXiv#1{\href {http://arxiv.org/abs/#1} {{\tt arXiv:#1}}}
\def\mn@eprint@dblp#1{\href {http://dblp.uni-trier.de/rec/bibtex/#1.xml}
  {dblp:#1}}
\def\mn@eprint@#1:#2:#3:#4\@nil{\def\@tempa {#1}\def\@tempb {#2}\def\@tempc
  {#3}\ifx \@tempc \@empty \let \@tempc \@tempb \let \@tempb \@tempa \fi \ifx
  \@tempb \@empty \def\@tempb {arXiv}\fi \@ifundefined
  {mn@eprint@\@tempb}{\@tempb:\@tempc}{\expandafter \expandafter \csname
  mn@eprint@\@tempb\endcsname \expandafter{\@tempc}}}

\bibitem[\protect\citeauthoryear{{Ag{\'u}ndez}, {Parmentier}, {Venot},
  {Hersant}  \& {Selsis}}{{Ag{\'u}ndez} et~al.}{2014a}]{2014A&A...564A..73A}
{Ag{\'u}ndez} M.,  {Parmentier} V.,  {Venot} O.,  {Hersant} F.,   {Selsis} F.,
  2014a, \mn@doi [\aap] {10.1051/0004-6361/201322895}, \href
  {https://ui.adsabs.harvard.edu/abs/2014A&A...564A..73A} {564, A73}

\bibitem[\protect\citeauthoryear{{Ag{\'u}ndez}, {Venot}, {Selsis}  \&
  {Iro}}{{Ag{\'u}ndez} et~al.}{2014b}]{2014ApJ...781...68A}
{Ag{\'u}ndez} M.,  {Venot} O.,  {Selsis} F.,   {Iro} N.,  2014b, \mn@doi [\apj]
  {10.1088/0004-637X/781/2/68}, \href
  {https://ui.adsabs.harvard.edu/abs/2014ApJ...781...68A} {781, 68}

\bibitem[\protect\citeauthoryear{{Allard}, {Homeier}, {Freytag}  \&
  {Sharp}}{{Allard} et~al.}{2012}]{Allard2012}
{Allard} F.,  {Homeier} D.,  {Freytag} B.,   {Sharp} C.~M.,  2012, in
  {Reyl{\'e}} C.,  {Charbonnel} C.,   {Schultheis} M.,  eds,  EAS Publications
  Series Vol. 57, EAS Publications Series. pp 3--43 (\mn@eprint {arXiv}
  {1206.1021}), \mn@doi{10.1051/eas/1257001}

\bibitem[\protect\citeauthoryear{{Amundsen}, {Baraffe}, {Tremblin}, {Manners},
  {Hayek}, {Mayne}  \& {Acreman}}{{Amundsen}
  et~al.}{2014}]{2014A&A...564A..59A}
{Amundsen} D.~S.,  {Baraffe} I.,  {Tremblin} P.,  {Manners} J.,  {Hayek} W.,
  {Mayne} N.~J.,   {Acreman} D.~M.,  2014, \mn@doi [\aap]
  {10.1051/0004-6361/201323169}, \href
  {https://ui.adsabs.harvard.edu/abs/2014A&A...564A..59A} {564, A59}

\bibitem[\protect\citeauthoryear{{Amundsen} et~al.,}{{Amundsen}
  et~al.}{2016}]{2016A&A...595A..36A}
{Amundsen} D.~S.,  et~al., 2016, \mn@doi [\aap] {10.1051/0004-6361/201629183},
  \href {https://ui.adsabs.harvard.edu/abs/2016A&A...595A..36A} {595, A36}

\bibitem[\protect\citeauthoryear{{Amundsen}, {Tremblin}, {Manners}, {Baraffe}
  \& {Mayne}}{{Amundsen} et~al.}{2017}]{AmuTM17}
{Amundsen} D.~S.,  {Tremblin} P.,  {Manners} J.,  {Baraffe} I.,   {Mayne}
  N.~J.,  2017, \mn@doi [\aap] {10.1051/0004-6361/201629322}, \href
  {http://adsabs.harvard.edu/abs/2017A%26A...598A..97A} {598, A97}

\bibitem[\protect\citeauthoryear{{Barber}, {Tennyson}, {Harris}  \&
  {Tolchenov}}{{Barber} et~al.}{2006}]{2006MNRAS.368.1087B}
{Barber} R.~J.,  {Tennyson} J.,  {Harris} G.~J.,   {Tolchenov} R.~N.,  2006,
  \mn@doi [\mnras] {10.1111/j.1365-2966.2006.10184.x}, \href
  {https://ui.adsabs.harvard.edu/abs/2006MNRAS.368.1087B} {368, 1087}

\bibitem[\protect\citeauthoryear{{Baudino}, {Molli{\`e}re}, {Venot},
  {Tremblin}, {B{\'e}zard}  \& {Lagage}}{{Baudino} et~al.}{2017}]{Baudino2017}
{Baudino} J.-L.,  {Molli{\`e}re} P.,  {Venot} O.,  {Tremblin} P.,  {B{\'e}zard}
  B.,   {Lagage} P.-O.,  2017, \mn@doi [\apj] {10.3847/1538-4357/aa95be}, \href
  {https://ui.adsabs.harvard.edu/abs/2017ApJ...850..150B} {850, 150}

\bibitem[\protect\citeauthoryear{{Bell} \& {Cowan}}{{Bell} \&
  {Cowan}}{2018}]{2018ApJ...857L..20B}
{Bell} T.~J.,  {Cowan} N.~B.,  2018, \mn@doi [\apjl]
  {10.3847/2041-8213/aabcc8}, \href
  {https://ui.adsabs.harvard.edu/abs/2018ApJ...857L..20B} {857, L20}

\bibitem[\protect\citeauthoryear{{Bell} et~al.,}{{Bell}
  et~al.}{2019}]{2019MNRAS.489.1995B}
{Bell} T.~J.,  et~al., 2019, \mn@doi [\mnras] {10.1093/mnras/stz2018}, \href
  {https://ui.adsabs.harvard.edu/abs/2019MNRAS.489.1995B} {489, 1995}

\bibitem[\protect\citeauthoryear{{Chakrabarty} \& {Sengupta}}{{Chakrabarty} \&
  {Sengupta}}{2019}]{2019AJ....158...39C}
{Chakrabarty} A.,  {Sengupta} S.,  2019, \mn@doi [\aj]
  {10.3847/1538-3881/ab24dd}, \href
  {https://ui.adsabs.harvard.edu/abs/2019AJ....158...39C} {158, 39}

\bibitem[\protect\citeauthoryear{{Cowan}, {Machalek}, {Croll}, {Shekhtman},
  {Burrows}, {Deming}, {Greene}  \& {Hora}}{{Cowan}
  et~al.}{2012}]{2012ApJ...747...82C}
{Cowan} N.~B.,  {Machalek} P.,  {Croll} B.,  {Shekhtman} L.~M.,  {Burrows} A.,
  {Deming} D.,  {Greene} T.,   {Hora} J.~L.,  2012, \mn@doi [\apj]
  {10.1088/0004-637X/747/1/82}, \href
  {https://ui.adsabs.harvard.edu/abs/2012ApJ...747...82C} {747, 82}

\bibitem[\protect\citeauthoryear{{D{\'e}sert}, {Vidal-Madjar}, {Lecavelier Des
  Etangs}, {Sing}, {Ehrenreich}, {H{\'e}brard}  \& {Ferlet}}{{D{\'e}sert}
  et~al.}{2008}]{2008A&A...492..585D}
{D{\'e}sert} J.~M.,  {Vidal-Madjar} A.,  {Lecavelier Des Etangs} A.,  {Sing}
  D.,  {Ehrenreich} D.,  {H{\'e}brard} G.,   {Ferlet} R.,  2008, \mn@doi [\aap]
  {10.1051/0004-6361:200810355}, \href
  {https://ui.adsabs.harvard.edu/abs/2008A&A...492..585D} {492, 585}

\bibitem[\protect\citeauthoryear{{Drummond}, {Tremblin}, {Baraffe}, {Amundsen},
  {Mayne}, {Venot}  \& {Goyal}}{{Drummond} et~al.}{2016}]{2016A&A...594A..69D}
{Drummond} B.,  {Tremblin} P.,  {Baraffe} I.,  {Amundsen} D.~S.,  {Mayne}
  N.~J.,  {Venot} O.,   {Goyal} J.,  2016, \mn@doi [\aap]
  {10.1051/0004-6361/201628799}, \href
  {https://ui.adsabs.harvard.edu/abs/2016A&A...594A..69D} {594, A69}

\bibitem[\protect\citeauthoryear{{Drummond}, {Mayne}, {Baraffe}, {Tremblin},
  {Manners}, {Amundsen}, {Goyal}  \& {Acreman}}{{Drummond}
  et~al.}{2018a}]{2018A&A...612A.105D}
{Drummond} B.,  {Mayne} N.~J.,  {Baraffe} I.,  {Tremblin} P.,  {Manners} J.,
  {Amundsen} D.~S.,  {Goyal} J.,   {Acreman} D.,  2018a, \mn@doi [\aap]
  {10.1051/0004-6361/201732010}, \href
  {https://ui.adsabs.harvard.edu/abs/2018A&A...612A.105D} {612, A105}

\bibitem[\protect\citeauthoryear{{Drummond} et~al.,}{{Drummond}
  et~al.}{2018b}]{2018ApJ...855L..31D}
{Drummond} B.,  et~al., 2018b, \mn@doi [\apjl] {10.3847/2041-8213/aab209},
  \href {https://ui.adsabs.harvard.edu/abs/2018ApJ...855L..31D} {855, L31}

\bibitem[\protect\citeauthoryear{{Drummond}, {Carter}, {H{\'e}brard}, {Mayne},
  {Sing}, {Evans}  \& {Goyal}}{{Drummond} et~al.}{2019}]{2019MNRAS.486.1123D}
{Drummond} B.,  {Carter} A.~L.,  {H{\'e}brard} E.,  {Mayne} N.~J.,  {Sing}
  D.~K.,  {Evans} T.~M.,   {Goyal} J.,  2019, \mn@doi [\mnras]
  {10.1093/mnras/stz909}, \href
  {https://ui.adsabs.harvard.edu/abs/2019MNRAS.486.1123D} {486, 1123}

\bibitem[\protect\citeauthoryear{{Drummond} et~al.,}{{Drummond}
  et~al.}{2020}]{2020A&A...636A..68D}
{Drummond} B.,  et~al., 2020, \mn@doi [\aap] {10.1051/0004-6361/201937153},
  \href {https://ui.adsabs.harvard.edu/abs/2020A&A...636A..68D} {636, A68}

\bibitem[\protect\citeauthoryear{{Evans} et~al.,}{{Evans}
  et~al.}{2017}]{Evans2017}
{Evans} T.~M.,  et~al., 2017, \mn@doi [\nat] {10.1038/nature23266}, \href
  {https://ui.adsabs.harvard.edu/abs/2017Natur.548...58E} {548, 58}

\bibitem[\protect\citeauthoryear{Fortney, Lodders, Marley  \& Freedman}{Fortney
  et~al.}{2008}]{Fortney_2008}
Fortney J.~J.,  Lodders K.,  Marley M.~S.,   Freedman R.~S.,  2008, \mn@doi
  [The Astrophysical Journal] {10.1086/528370}, 678, 1419

\bibitem[\protect\citeauthoryear{Gandhi \& Madhusudhan}{Gandhi \&
  Madhusudhan}{2017}]{Gandhi_2017}
Gandhi S.,  Madhusudhan N.,  2017, \mn@doi [Monthly Notices of the Royal
  Astronomical Society] {10.1093/mnras/stx1601}, 472, 2334

\bibitem[\protect\citeauthoryear{{Gordon} \& {McBride}}{{Gordon} \&
  {McBride}}{1994}]{GordonMcBride1994}
{Gordon} S.,  {McBride} B.~J.,  1994, NASA Reference Publication, 1311

\bibitem[\protect\citeauthoryear{{Goyal} et~al.,}{{Goyal}
  et~al.}{2018}]{2018MNRAS.474.5158G}
{Goyal} J.~M.,  et~al., 2018, \mn@doi [\mnras] {10.1093/mnras/stx3015}, \href
  {https://ui.adsabs.harvard.edu/abs/2018MNRAS.474.5158G} {474, 5158}

\bibitem[\protect\citeauthoryear{{Goyal} et~al.,}{{Goyal}
  et~al.}{2020}]{2020MNRAS.tmp.2424G}
{Goyal} J.~M.,  et~al., 2020, \mn@doi [\mnras] {10.1093/mnras/staa2300}, \href
  {https://ui.adsabs.harvard.edu/abs/2020MNRAS.tmp.2424G} {}

\bibitem[\protect\citeauthoryear{{Guillot} \& {Showman}}{{Guillot} \&
  {Showman}}{2002}]{2002A&A...385..156G}
{Guillot} T.,  {Showman} A.~P.,  2002, \mn@doi [\aap]
  {10.1051/0004-6361:20011624}, \href
  {https://ui.adsabs.harvard.edu/abs/2002A&A...385..156G} {385, 156}

\bibitem[\protect\citeauthoryear{{Heiter} et~al.,}{{Heiter}
  et~al.}{2015}]{2015PhyS...90e4010H}
{Heiter} U.,  et~al., 2015, \mn@doi [\physscr] {10.1088/0031-8949/90/5/054010},
  \href {https://ui.adsabs.harvard.edu/abs/2015PhyS...90e4010H} {90, 054010}

\bibitem[\protect\citeauthoryear{{Helling}}{{Helling}}{2020}]{2020arXiv201103302H}
{Helling} C.,  2020, arXiv e-prints, \href
  {https://ui.adsabs.harvard.edu/abs/2020arXiv201103302H} {p. arXiv:2011.03302}

\bibitem[\protect\citeauthoryear{{Iro}, {B{\'e}zard}  \& {Guillot}}{{Iro}
  et~al.}{2005}]{2005A&A...436..719I}
{Iro} N.,  {B{\'e}zard} B.,   {Guillot} T.,  2005, \mn@doi [\aap]
  {10.1051/0004-6361:20048344}, \href
  {https://ui.adsabs.harvard.edu/abs/2005A&A...436..719I} {436, 719}

\bibitem[\protect\citeauthoryear{{John}}{{John}}{1988}]{John1988}
{John} T.~L.,  1988, \aap, \href
  {https://ui.adsabs.harvard.edu/abs/1988A&A...193..189J} {193, 189}

\bibitem[\protect\citeauthoryear{Komacek \& Tan}{Komacek \&
  Tan}{2018}]{Komacek_2018}
Komacek T.~D.,  Tan X.,  2018, \mn@doi [Research Notes of the {AAS}]
  {10.3847/2515-5172/aac5e7}, 2, 36

\bibitem[\protect\citeauthoryear{{Komacek}, {Showman}  \& {Tan}}{{Komacek}
  et~al.}{2017}]{2017ApJ...835..198K}
{Komacek} T.~D.,  {Showman} A.~P.,   {Tan} X.,  2017, \mn@doi [\apj]
  {10.3847/1538-4357/835/2/198}, \href
  {https://ui.adsabs.harvard.edu/abs/2017ApJ...835..198K} {835, 198}

\bibitem[\protect\citeauthoryear{{Lacis} \& {Oinas}}{{Lacis} \&
  {Oinas}}{1991}]{Lacis1991}
{Lacis} A.~A.,  {Oinas} V.,  1991, \mn@doi [\jgr] {10.1029/90JD01945}, \href
  {http://adsabs.harvard.edu/abs/1991JGR....96.9027L} {96, 9027}

\bibitem[\protect\citeauthoryear{{Malik} et~al.,}{{Malik}
  et~al.}{2017}]{2017AJ....153...56M}
{Malik} M.,  et~al., 2017, \mn@doi [\aj] {10.3847/1538-3881/153/2/56}, \href
  {https://ui.adsabs.harvard.edu/abs/2017AJ....153...56M} {153, 56}

\bibitem[\protect\citeauthoryear{{Malik}, {Kitzmann}, {Mendon{\c{c}}a},
  {Grimm}, {Marleau}, {Linder}, {Tsai}  \& {Heng}}{{Malik}
  et~al.}{2019}]{Malik2019}
{Malik} M.,  {Kitzmann} D.,  {Mendon{\c{c}}a} J.~M.,  {Grimm} S.~L.,  {Marleau}
  G.-D.,  {Linder} E.~F.,  {Tsai} S.-M.,   {Heng} K.,  2019, \mn@doi [\aj]
  {10.3847/1538-3881/ab1084}, \href
  {https://ui.adsabs.harvard.edu/abs/2019AJ....157..170M} {157, 170}

\bibitem[\protect\citeauthoryear{Mansfield et~al.,}{Mansfield
  et~al.}{2020}]{Mansfield_2020}
Mansfield M.,  et~al., 2020, \mn@doi [The Astrophysical Journal]
  {10.3847/2041-8213/ab5b09}, 888, L15

\bibitem[\protect\citeauthoryear{{Mayne} et~al.,}{{Mayne}
  et~al.}{2014}]{2014A&A...561A...1M}
{Mayne} N.~J.,  et~al., 2014, \mn@doi [\aap] {10.1051/0004-6361/201322174},
  \href {https://ui.adsabs.harvard.edu/abs/2014A&A...561A...1M} {561, A1}

\bibitem[\protect\citeauthoryear{Mayne et~al.,}{Mayne et~al.}{2017}]{Mayne2017}
Mayne N.~J.,  et~al., 2017, \mn@doi [Astronomy and Astrophysics]
  {10.1051/0004-6361/201730465}, 604, 1

\bibitem[\protect\citeauthoryear{{McBride}, {Gordon}  \& {Reno}}{{McBride}
  et~al.}{1993}]{Mcbride1993}
{McBride} B.~J.,  {Gordon} S.,   {Reno} M.~A.,  1993, NASA Technical
  Memorandum, 4513

\bibitem[\protect\citeauthoryear{{McBride}, {Zehe}  \& {Gordon}}{{McBride}
  et~al.}{2002}]{McBride2002}
{McBride} B.~J.,  {Zehe} M.~J.,   {Gordon} S.,  2002, NASA/TP, 2002-211556

\bibitem[\protect\citeauthoryear{{McKemmish}, {Yurchenko}  \&
  {Tennyson}}{{McKemmish} et~al.}{2016}]{2016MNRAS.463..771M}
{McKemmish} L.~K.,  {Yurchenko} S.~N.,   {Tennyson} J.,  2016, \mn@doi [\mnras]
  {10.1093/mnras/stw1969}, \href
  {https://ui.adsabs.harvard.edu/abs/2016MNRAS.463..771M} {463, 771}

\bibitem[\protect\citeauthoryear{{Merritt} et~al.,}{{Merritt}
  et~al.}{2020}]{Merritt_2020}
{Merritt} S.~R.,  et~al., 2020, \mn@doi [A\&A] {10.1051/0004-6361/201937409},
  636, A117

\bibitem[\protect\citeauthoryear{{Mikal-Evans}, {Sing}, {Kataria}, {Wakeford},
  {Mayne}, {Lewis}, {Barstow}  \& {Spake}}{{Mikal-Evans}
  et~al.}{2020}]{Mikal-Evans2020}
{Mikal-Evans} T.,  {Sing} D.~K.,  {Kataria} T.,  {Wakeford} H.~R.,  {Mayne}
  N.~J.,  {Lewis} N.~K.,  {Barstow} J.~K.,   {Spake} J.~J.,  2020, \mn@doi
  [\mnras] {10.1093/mnras/staa1628}, \href
  {https://ui.adsabs.harvard.edu/abs/2020MNRAS.496.1638M} {496, 1638}

\bibitem[\protect\citeauthoryear{Molli{\`{e}}re, van Boekel, Dullemond, Henning
   \& Mordasini}{Molli{\`{e}}re et~al.}{2015}]{Molliere_2015}
Molli{\`{e}}re P.,  van Boekel R.,  Dullemond C.,  Henning T.,   Mordasini C.,
  2015, \mn@doi [The Astrophysical Journal] {10.1088/0004-637x/813/1/47}, 813,
  47

\bibitem[\protect\citeauthoryear{{Moses}, {Madhusudhan}, {Visscher}  \&
  {Freedman}}{{Moses} et~al.}{2013}]{2013ApJ...763...25M}
{Moses} J.~I.,  {Madhusudhan} N.,  {Visscher} C.,   {Freedman} R.~S.,  2013,
  \mn@doi [\apj] {10.1088/0004-637X/763/1/25}, \href
  {https://ui.adsabs.harvard.edu/abs/2013ApJ...763...25M} {763, 25}

\bibitem[\protect\citeauthoryear{{Nikolov} et~al.,}{{Nikolov}
  et~al.}{2018}]{Nikolov2018}
{Nikolov} N.,  et~al., 2018, \mn@doi [\mnras] {10.1093/mnras/stx2865}, \href
  {https://ui.adsabs.harvard.edu/abs/2018MNRAS.474.1705N} {474, 1705}

\bibitem[\protect\citeauthoryear{{Parmentier}, {Showman}  \&
  {Lian}}{{Parmentier} et~al.}{2013}]{Parmentier2013}
{Parmentier} V.,  {Showman} A.~P.,   {Lian} Y.,  2013, \mn@doi [\aap]
  {10.1051/0004-6361/201321132}, \href
  {https://ui.adsabs.harvard.edu/abs/2013A&A...558A..91P} {558, A91}

\bibitem[\protect\citeauthoryear{{Parmentier} et~al.,}{{Parmentier}
  et~al.}{2018}]{2018A&A...617A.110P}
{Parmentier} V.,  et~al., 2018, \mn@doi [\aap] {10.1051/0004-6361/201833059},
  \href {https://ui.adsabs.harvard.edu/abs/2018A&A...617A.110P} {617, A110}

\bibitem[\protect\citeauthoryear{{Parmentier}, {Showman}  \&
  {Fortney}}{{Parmentier} et~al.}{2020}]{2020MNRAS.tmp.3310P}
{Parmentier} V.,  {Showman} A.~P.,   {Fortney} J.~J.,  2020, \mn@doi [\mnras]
  {10.1093/mnras/staa3418}, \href
  {https://ui.adsabs.harvard.edu/abs/2020MNRAS.tmp.3310P} {}

\bibitem[\protect\citeauthoryear{{Phillips} et~al.,}{{Phillips}
  et~al.}{2020}]{2020arXiv200313717P}
{Phillips} M.~W.,  et~al., 2020, arXiv e-prints, \href
  {https://ui.adsabs.harvard.edu/abs/2020arXiv200313717P} {p. arXiv:2003.13717}

\bibitem[\protect\citeauthoryear{{Plez}}{{Plez}}{1998}]{Plez1998}
{Plez} B.,  1998, \aap, \href
  {https://ui.adsabs.harvard.edu/abs/1998A&A...337..495P} {337, 495}

\bibitem[\protect\citeauthoryear{{Richard} et~al.,}{{Richard}
  et~al.}{2012}]{2012JQSRT.113.1276R}
{Richard} C.,  et~al., 2012, \mn@doi [\jqsrt] {10.1016/j.jqsrt.2011.11.004},
  \href {https://ui.adsabs.harvard.edu/abs/2012JQSRT.113.1276R} {113, 1276}

\bibitem[\protect\citeauthoryear{{Rogers} \& {Komacek}}{{Rogers} \&
  {Komacek}}{2014}]{2014ApJ...794..132R}
{Rogers} T.~M.,  {Komacek} T.~D.,  2014, \mn@doi [\apj]
  {10.1088/0004-637X/794/2/132}, \href
  {https://ui.adsabs.harvard.edu/abs/2014ApJ...794..132R} {794, 132}

\bibitem[\protect\citeauthoryear{{Rothman} et~al.,}{{Rothman}
  et~al.}{2010}]{2010JQSRT.111.2139R}
{Rothman} L.~S.,  et~al., 2010, \mn@doi [\jqsrt] {10.1016/j.jqsrt.2010.05.001},
  \href {https://ui.adsabs.harvard.edu/abs/2010JQSRT.111.2139R} {111, 2139}

\bibitem[\protect\citeauthoryear{Ryabchikova, Piskunov, Kurucz, Stempels,
  Heiter, Pakhomov  \& Barklem}{Ryabchikova et~al.}{2015}]{Ryabchikova_2015}
Ryabchikova T.,  Piskunov N.,  Kurucz R.~L.,  Stempels H.~C.,  Heiter U.,
  Pakhomov Y.,   Barklem P.~S.,  2015, \mn@doi [Physica Scripta]
  {10.1088/0031-8949/90/5/054005}, 90, 054005

\bibitem[\protect\citeauthoryear{{Snellen}, {de Kok}, {de Mooij}  \&
  {Albrecht}}{{Snellen} et~al.}{2010}]{2010Natur.465.1049S}
{Snellen} I. A.~G.,  {de Kok} R.~J.,  {de Mooij} E. J.~W.,   {Albrecht} S.,
  2010, \mn@doi [\nat] {10.1038/nature09111}, \href
  {https://ui.adsabs.harvard.edu/abs/2010Natur.465.1049S} {465, 1049}

\bibitem[\protect\citeauthoryear{Swain et~al.,}{Swain
  et~al.}{2013}]{Swain_2013}
Swain M.,  et~al., 2013, \mn@doi [Icarus]
  {https://doi.org/10.1016/j.icarus.2013.04.003}, 225, 432

\bibitem[\protect\citeauthoryear{{Tan} \& {Komacek}}{{Tan} \&
  {Komacek}}{2019}]{2019ApJ...886...26T}
{Tan} X.,  {Komacek} T.~D.,  2019, \mn@doi [\apj] {10.3847/1538-4357/ab4a76},
  \href {https://ui.adsabs.harvard.edu/abs/2019ApJ...886...26T} {886, 26}

\bibitem[\protect\citeauthoryear{{Tashkun} \& {Perevalov}}{{Tashkun} \&
  {Perevalov}}{2011}]{2011JQSRT.112.1403T}
{Tashkun} S.~A.,  {Perevalov} V.~I.,  2011, \mn@doi [\jqsrt]
  {10.1016/j.jqsrt.2011.03.005}, \href
  {https://ui.adsabs.harvard.edu/abs/2011JQSRT.112.1403T} {112, 1403}

\bibitem[\protect\citeauthoryear{{Tremblin}, {Amundsen}, {Mourier}, {Baraffe},
  {Chabrier}, {Drummond}, {Homeier}  \& {Venot}}{{Tremblin}
  et~al.}{2015}]{2015ApJ...804L..17T}
{Tremblin} P.,  {Amundsen} D.~S.,  {Mourier} P.,  {Baraffe} I.,  {Chabrier} G.,
   {Drummond} B.,  {Homeier} D.,   {Venot} O.,  2015, \mn@doi [\apjl]
  {10.1088/2041-8205/804/1/L17}, \href
  {https://ui.adsabs.harvard.edu/abs/2015ApJ...804L..17T} {804, L17}

\bibitem[\protect\citeauthoryear{{Tremblin}, {Amundsen}, {Chabrier}, {Baraffe},
  {Drummond}, {Hinkley}, {Mourier}  \& {Venot}}{{Tremblin}
  et~al.}{2016}]{2016ApJ...817L..19T}
{Tremblin} P.,  {Amundsen} D.~S.,  {Chabrier} G.,  {Baraffe} I.,  {Drummond}
  B.,  {Hinkley} S.,  {Mourier} P.,   {Venot} O.,  2016, \mn@doi [\apjl]
  {10.3847/2041-8205/817/2/L19}, \href
  {https://ui.adsabs.harvard.edu/abs/2016ApJ...817L..19T} {817, L19}

\bibitem[\protect\citeauthoryear{{Tremblin} et~al.,}{{Tremblin}
  et~al.}{2017}]{2017ApJ...841...30T}
{Tremblin} P.,  et~al., 2017, \mn@doi [\apj] {10.3847/1538-4357/aa6e57}, \href
  {https://ui.adsabs.harvard.edu/abs/2017ApJ...841...30T} {841, 30}

\bibitem[\protect\citeauthoryear{{Wakeford} et~al.,}{{Wakeford}
  et~al.}{2018}]{Wakeford2018}
{Wakeford} H.~R.,  et~al., 2018, \mn@doi [\aj] {10.3847/1538-3881/aa9e4e},
  \href {https://ui.adsabs.harvard.edu/abs/2018AJ....155...29W} {155, 29}

\bibitem[\protect\citeauthoryear{{Yurchenko} \& {Tennyson}}{{Yurchenko} \&
  {Tennyson}}{2014}]{2014MNRAS.440.1649Y}
{Yurchenko} S.~N.,  {Tennyson} J.,  2014, \mn@doi [\mnras]
  {10.1093/mnras/stu326}, \href
  {https://ui.adsabs.harvard.edu/abs/2014MNRAS.440.1649Y} {440, 1649}

\bibitem[\protect\citeauthoryear{{Yurchenko}, {Barber}  \&
  {Tennyson}}{{Yurchenko} et~al.}{2011}]{2011MNRAS.413.1828Y}
{Yurchenko} S.~N.,  {Barber} R.~J.,   {Tennyson} J.,  2011, \mn@doi [\mnras]
  {10.1111/j.1365-2966.2011.18261.x}, \href
  {https://ui.adsabs.harvard.edu/abs/2011MNRAS.413.1828Y} {413, 1828}

\makeatother
\end{thebibliography}

% Alternatively you could enter them by hand, like this:
% This method is tedious and prone to error if you have lots of references
%\begin{thebibliography}{99}
%\bibitem[\protect\citeauthoryear{Author}{2012}]{Author2012}
%Author A.~N., 2013, Journal of Improbable Astronomy, 1, 1
%\bibitem[\protect\citeauthoryear{Others}{2013}]{Others2013}
%Others S., 2012, Journal of Interesting Stuff, 17, 198
%\end{thebibliography}

%%%%%%%%%%%%%%%%%%%%%%%%%%%%%%%%%%%%%%%%%%%%%%%%%%

%%%%%%%%%%%%%%%%% APPENDICES %%%%%%%%%%%%%%%%%%%%%
%
\appendix
\section{Testing the Time-Dependent Radiative Equilibrium Solver}
\label{appendix:benchmark}
\begin{figure}
\centering 
\includegraphics[width=0.5\textwidth]{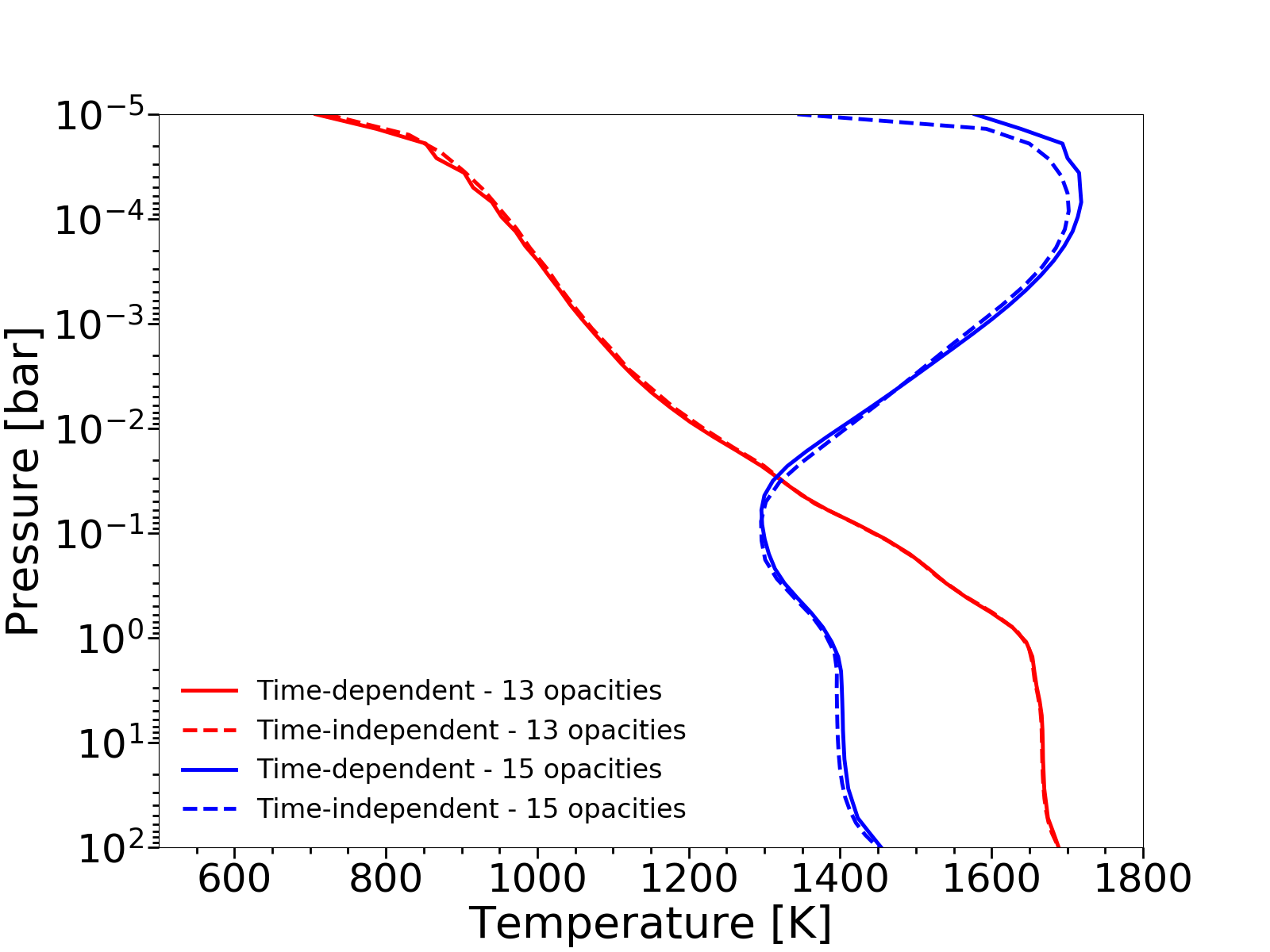} 
\caption{Comparison between the steady-state $P$ -- $T$ profiles from the time-independent 1D model (dashed lines) and the time-dependent 1D model (solid lines) for the model parameters used in \cref{subsec:tdmodel2}, with TiO/VO either excluded (red lines) or included (blue lines) as atmospheric opacity sources}
\label{fig:benchmark}
\end{figure}

The new time-dependent 1D model is benchmarked against a calculation using the well-establish time-independent method implemented in ATMO for HD209458b (see \cref{fig:benchmark}), to check for consistency. Parameters used can be seen under HD209458b in \cref{tab:params}. A heat redistribution factor of 0.5 and daytime average zenith angle of 60$^{\circ}$ were also used for the time-dependent 1D model. These tests are done both with and without TiO and VO included as opacity sources. These two molecules are strong UV/optical stellar flux absorbers, causing a temperature inversion in the atmosphere of HD209458b and many other hot Jupiters \citep{2008A&A...492..585D}. As expected, the additional molecules put more strain on the radiative solver, causing some minor deviations to appear at higher altitudes. Overall, however, there is a remarkable degree of agreement between the two approaches, given they use separate methods to determine atmospheric $P$ -- $T$ structure. The time-independent radiative-equilibrium solver has previously been compared against other 1D atmosphere models with generally very good agreement \citep[see][]{2016A&A...594A..69D, Baudino2017, Malik2019}.
\section{The Heat Capacity of a Mixture}
\label{appendix:heat_capacity}
The heat capacity $C_P$ of a substance at constant pressure is defined as
\begin{ceqn}
\begin{equation}
C_{P} = \left(\frac{\partial H}{\partial T}\right)_{P}=\left(\frac{\partial Q}{\partial T}\right)_{P}.
\end{equation}
\end{ceqn}
Here $C_P$  is the molar heat capacity with units J mol$^{-1}$ K$^{-1}$. The \textit{specific} heat capacity, $c_P$ , or the heat capacity per unit mass, has units J kg$^{-1}$ K$^{-1}$. The specific and molar heat capacities are \textit{intensive} properties of the substance. By contrast, the \textit{extensive} form of heat capacity $C$, measured in [J K$^{-1}$], is related to the intensive forms via
\begin{ceqn}
\begin{equation}
\begin{aligned}
c_{P} &= \left(\frac{\partial C}{\partial m}\right)_{P}, \\
C_{P} &= \left(\frac{\partial C}{\partial n}\right)_{P},
\end{aligned}
\end{equation}
\end{ceqn}
where $m$ is the mass of the substance, and $n$ is the total number of moles.

Following \citet{GordonMcBride1994}, we now derive the specific heat capacity of a mixture at chemical equilibrium. We start with the total enthalpy of a mixture, which is given by
\begin{ceqn}
\begin{equation}
h=\sum_{j} n_{j} H_{j}^{0},
\end{equation}
\end{ceqn}
\noindent
where $h$ is the total specific enthalpy of the system in J kg$^{-1}$, $n_j$ is the number of moles per kilogram of species $j$ in mol kg$^{-1}$ and $H_j^0$ is the molar enthalpy of the species $j$ in J mol$^{-1}$.
Now, using the definition of the heat capacity, we find the total \textit{specific} heat capacity of the mixture:
\begin{ceqn}
\begin{equation}
\begin{aligned}
c_{P} &=\left(\frac{\partial h}{\partial T}\right)_{P}, \\
c_{P} &=\sum_{j} n_{j}\left(\frac{\partial H_{j}^{0}}{\partial T}\right)_{p}+\sum_{j} H_{j}^{0}\left(\frac{\partial n_{j}}{\partial T}\right)_{p}, \\
c_{P} &=\sum_{j} n_{j} C_{P}^{0}+\sum_{j} H_{j}^{0}\left(\frac{\partial n_{j}}{\partial T}\right)_{P},
\end{aligned}
\end{equation}
\end{ceqn}
where $C_P^0$ is the molar standard heat capacity of the species $j$. 

It is clear that the total specific heat capacity of the mixture is a combination of two parts. The first term on the right in the final equation represents the sum of the molar heat capacities of the individual species (multiplied by the number of moles per kilogram, in this case). This component is referred to as the `frozen' heat capacity by \citet{GordonMcBride1994}. The second term, called the `reaction' heat capacity, contains a derivative of the number of moles (per unit mass) with temperature. This term becomes important when the abundance of a species varies rapidly with the temperature, for instance at a pressure-temperature region where the chemistry transitions from a \ch{CO}-dominated to a \ch{CH4}-dominated atmosphere, or in the region of a condensation curve.
The molar heat capacity can be easily obtained by multiplying the specific heat capacity by the mean molecular weight of the substance
\begin{ceqn}
\begin{equation}
C_P=\mu c_P
\end{equation}
\end{ceqn}
where $\mu$ is in kg mol$^{-1}$.
\section{Effect of Planetary Mass/Surface Gravity}
\label{appendix:grav}
\begin{figure*}
\centering 
\includegraphics[width=\textwidth]{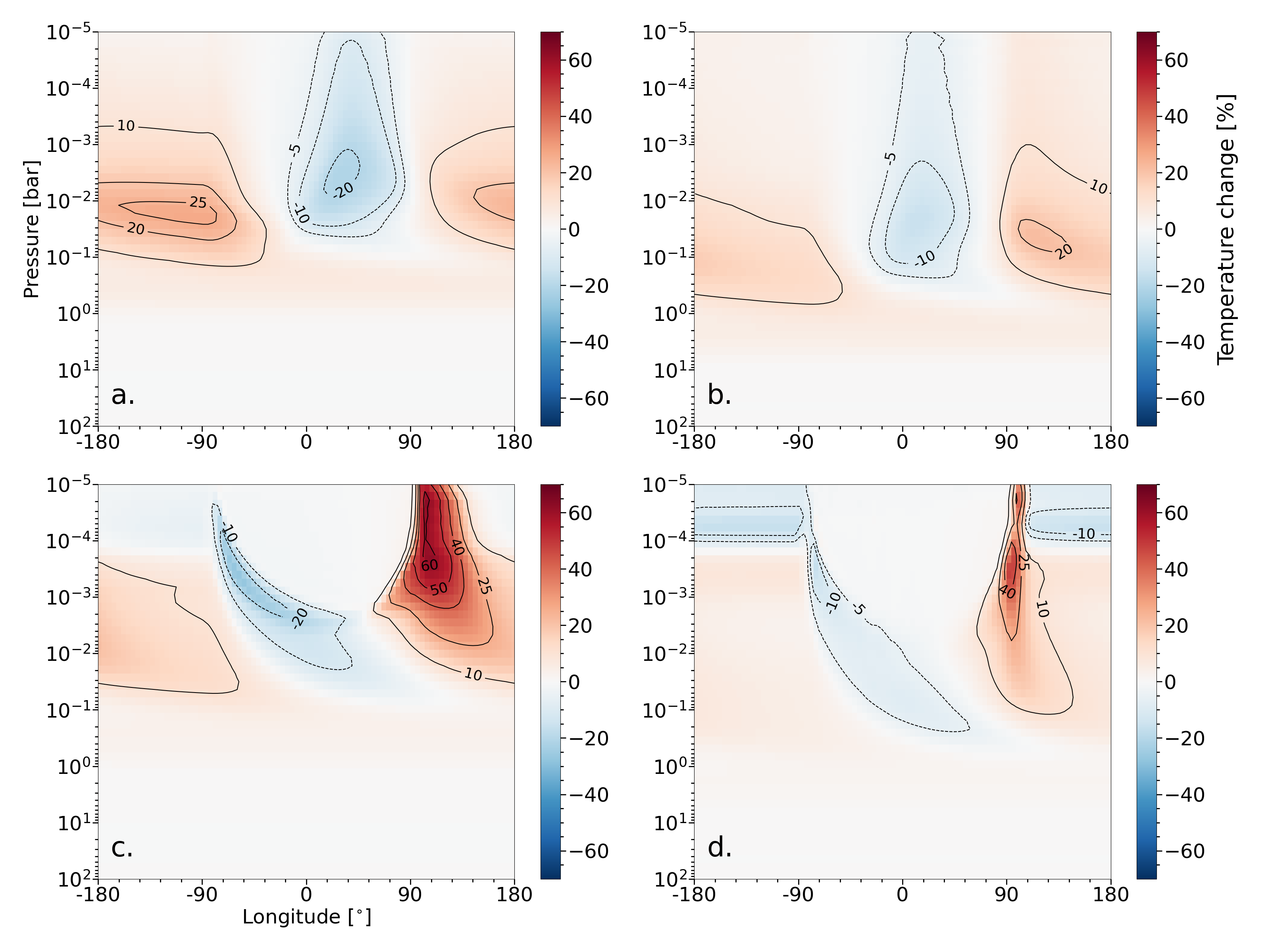}
\caption{Percentage change in temperature when including the reaction heat capacity with \ch{TiO}/\ch{VO} either excluded (top) or included (bottom) as atmospheric opacity sources and with a planetary mass of 0.32 $M_J$ ($\log(g)=2.36$) (left) or 1.47 $M_J$ ($\log(g)=3.02$) (right), respectively. Wind speed is set at 5 km s$^{-1}$.}
\label{fig:figA2}
\end{figure*}

In order to briefly investigate the effect of planetary mass, and corresponding surface gravity, on the reaction heat redistribution we ran the model selecting a much lower planetary mass of 0.32 $M_J$ ($\log(g)$=2.36). A comparison between the heat redistribution in this scenario and a much higher mass of 1.47 $M_J$ ($\log(g)$=3.02) can be seen in \cref{fig:figA2}. In the lower mass case, the impact of the reaction heat redistribution is significantly increased both with or without \ch{TiO} and \ch{VO}. This is because lower surface gravity increases the radiative timescale (\cref{eq:eq2}), causing the radiative heating and cooling of the atmosphere to have a decreasing relative effect when compared to the reaction heat redistribution. As the deep atmosphere takes much longer to respond to the stellar irradiation, the primary locations of both dissociation and recombination are also seen to shift towards lower pressure levels. Protracted low pressure night side cooling is reduced in the lower mass case for the same reason. As the deep atmosphere experiences less radiative heating, the high altitude cooling required for energy balance is decreased, see \cref{subsec:ws} for more detail.
\section{Emission Spectra}
\label{appendix:spectra}
\begin{figure*}
\centering 
\includegraphics[width=\textwidth]{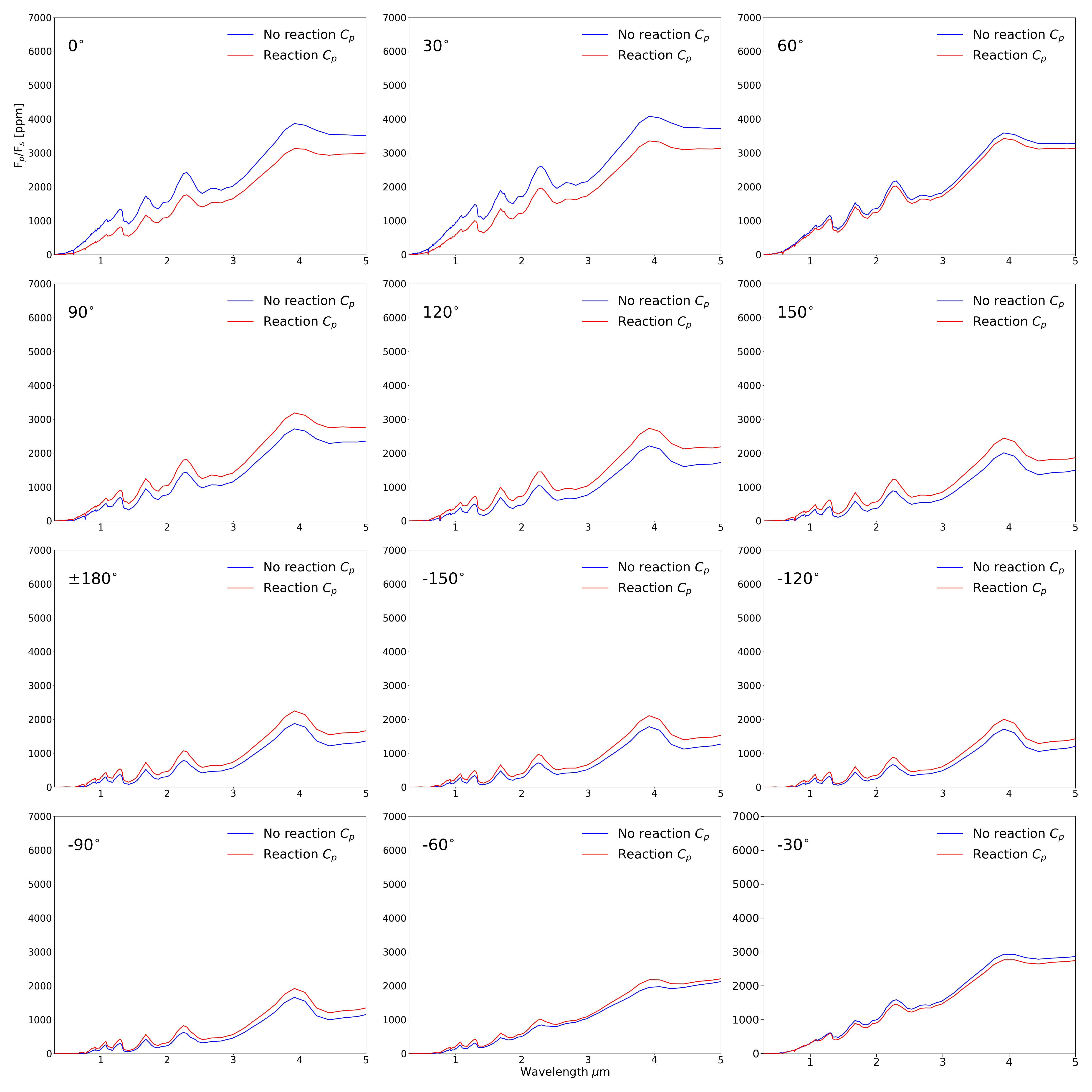}
\caption{Thermal emission spectra calculated from $P$ -- $T$ profiles situated every 30$^\circ$ longitude around the equator, for the models excluding \ch{TiO}/\ch{VO} as atmospheric opacity sources.}
\label{fig:figA3}
\end{figure*}

\begin{figure*}
\centering 
\includegraphics[width=\textwidth]{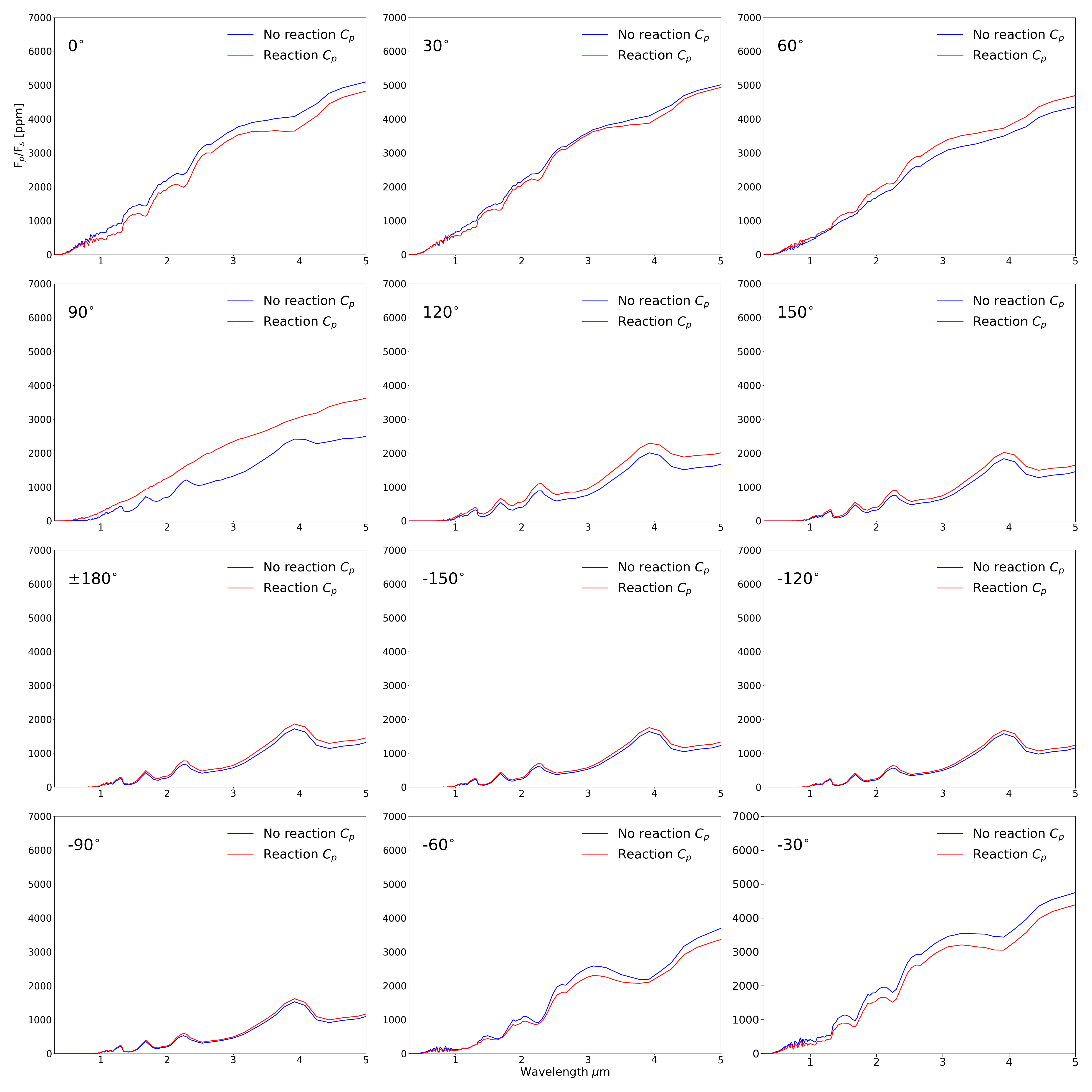}
\caption{Thermal emission spectra calculated from $P$ -- $T$ profiles situated every 30$^\circ$ longitude around the equator, for the models including \ch{TiO}/\ch{VO} as atmospheric opacity sources.}
\label{fig:figA4}
\end{figure*}

In order to construct the pseudo-2D phase curves seen in \cref{subsec:pc}, individual $P$ -- $T$ profiles are extracted from the steady-state $P$ -- $T$ structures at $5^{\circ}$ longitudinal intervals. These $P$ -- $T$ profiles are then used in ATMO's atmospheric emission routine, to calculate a series of emission spectra from which the pseudo-2D phase curves are constructed for wavelengths of 3.6~$\mu m$ and 4.5~$\mu m$. A sample of these emission spectra, excluding (\cref{fig:figA3}) and including (\cref{fig:figA4}) \ch{TiO}/\ch{VO}, ranging from 0.3-5 $\mu m$ at 30$^{\circ}$ intervals in longitude can be seen here.

%%%%%%%%%%%%%%%%%%%%%%%%%%%%%%%%%%%%%%%%%%%%%%%%%%

% Don't change these lines
\bsp	% typesetting comment
\label{lastpage}
\end{document}